\documentclass[article,longauth]{aa} 
\usepackage{graphicx}
\usepackage[colorlinks=true,linkcolor=blue,citecolor=blue]{hyperref} 

\usepackage{txfonts}
\usepackage{orcidlink}
\usepackage{xcolor}
\usepackage{float}
\usepackage{soul}

\begin{document} 

\title{Unveiling the HD 95086 system at mid-infrared wavelengths with JWST/MIRI}
\subtitle{}

\author{
    Mathilde Mâlin\orcidlink{0000-0002-2918-8479}\inst{\ref{lesia},\ref{stsci},\ref{jhu}},
    Anthony Boccaletti\orcidlink{0000-0001-9353-2724}\inst{\ref{lesia}}, 
    Clément Perrot\orcidlink{0000-0003-3831-0381}\inst{\ref{lesia}},
    Pierre Baudoz\orcidlink{0000-0002-2711-7116}\inst{\ref{lesia}}, 
    Daniel Rouan\orcidlink{0000-0002-2352-1736}\inst{\ref{lesia}},
    Pierre-Olivier Lagage\inst{\ref{cea}},
    Rens Waters\orcidlink{0000-0002-5462-9387}\inst{\ref{radboud},\ref{sron}},
    Manuel G\"udel\orcidlink{0000-0001-9818-0588}\inst{\ref{vienna},\ref{mpia},\ref{eth}},
    Thomas Henning\orcidlink{0000-0002-1493-300X}\inst{\ref{mpia}},
    Bart Vandenbussche\orcidlink{0000-0002-1368-3109}\inst{\ref{leuven}},
    Olivier Absil\orcidlink{0000-0002-4006-6237}\inst{\ref{star}},
    David Barrado\orcidlink{0000-0002-5971-9242}\inst{\ref{cab}},
    Christophe Cossou\orcidlink{0000-0001-5350-4796}\inst{\ref{parissaclay}}, 
    Leen Decin\orcidlink{0000-0002-5342-8612}\inst{\ref{leuven}},
    Adrian M. Glauser\orcidlink{0000-0001-9250-1547}\inst{\ref{eth}},
    John Pye\orcidlink{0000-0002-0932-4330}\inst{\ref{leicester}},
    Goran Olofsson\orcidlink{0000-0003-3747-7120}\inst{\ref{stockholm}},
    Alistair Glasse\orcidlink{0000-0002-2041-2462}\inst{\ref{ukatc}},
    Fred Lahuis\orcidlink{0000-0002-8443-9488}\inst{\ref{sron}},
    Polychronis Patapis\orcidlink{0000-0001-8718-3732}\inst{\ref{eth}},
    Pierre Royer\orcidlink{0000-0001-9341-2546}\inst{\ref{leuven}},
    Silvia Scheithauer\orcidlink{0000-0003-4559-0721}\inst{\ref{mpia}},
    Niall Whiteford\orcidlink{0000-0001-8818-1544}\inst{\ref{museum}},
    Eugene Serabyn\inst{\ref{jpl}},
    Elodie Choquet\orcidlink{0000-0002-9173-0740}\inst{\ref{lam}},
    Luis Colina\inst{\ref{cab}},
    G\"oran Ostlin\orcidlink{0000-0002-3005-1349}\inst{\ref{oskar}},
    Tom P.\ Ray\orcidlink{0000-0002-2110-1068}\inst{\ref{dublin}}
    Gillian Wright\orcidlink{0000-0001-7416-7936}\inst{\ref{ukatc}}
    }
\institute{
LESIA, Observatoire de Paris, Universit{\'e} PSL, CNRS, Sorbonne Universit{\'e}, Univ. Paris Diderot, Sorbonne Paris Cit{\'e}, 5 place Jules Janssen, 92195 Meudon, France\label{lesia} 
\and Space Telescope Science Institute, 3700 San Martin Drive, Baltimore, MD 21218, USA\label{stsci}
\and Department of Physics \& Astronomy, Johns Hopkins University, 3400 N. Charles Street, Baltimore, MD 21218, USA\label{jhu}
\and Universit{\'e} Paris-Saclay, Universit{\'e} Paris Cit{\'e}, CEA, CNRS, AIM, 91191, Gif-sur-Yvette, France\label{cea}
\and  Department of Astrophysics/IMAPP, Radboud University, PO Box 9010, 6500 GL Nijmegen, the Netherlands\label{radboud}
\and  SRON Netherlands Institute for Space Research, Niels Bohrweg 4, 2333 CA Leiden, the Netherlands\label{sron}
\and   Department of Astrophysics, University of Vienna, T\"urkenschanzstrasse 17, 1180 Vienna, Austria\label{vienna}
\and  Max-Planck-Institut f\"ur Astronomie (MPIA), K\"onigstuhl 17, 69117 Heidelberg, Germany \label{mpia}
\and ETH Z\"urich, Institute for Particle Physics and Astrophysics, Wolfgang-Pauli-Strasse 27, 8093 Z\"urich, Switzerland\label{eth}
\and Institute of Astronomy, KU Leuven, Celestijnenlaan 200D, 3001 Leuven, Belgium\label{leuven}
\and STAR Institute, Universit\'e de Li\`ege, All\'ee du Six Ao\^ut 19c, 4000 Li\`ege, Belgium\label{star}
\and Centro de Astrobiología (CAB), CSIC-INTA, ESAC Campus, Camino Bajo del Castillo s/n, 28692 Villanueva de la Cañada, Madrid, Spain \label{cab}
\and  School of Physics \& Astronomy, Space Park Leicester, University of Leicester, 92 Corporation Road, Leicester, LE4 5SP, UK\label{leicester}
\and Department of Astronomy, Stockholm University, AlbaNova University Center, 10691 Stockholm, Sweden\label{stockholm}
\and UK Astronomy Technology Centre, Royal Observatory, Blackford Hill, Edinburgh EH9 3HJ, UK\label{ukatc}
\and Université Paris-Saclay, CEA, IRFU, 91191, Gif-sur-Yvette, France\label{parissaclay}
\and Department of Astrophysics, American Museum of Natural History, New York, NY 10024, USA\label{museum}
\and Jet Propulsion Laboratory, California Institute of Technology, 4800 Oak Grove Dr.,Pasadena, CA 91109, USA\label{jpl}
\and Aix Marseille Univ, CNRS, CNES, LAM, Marseille, France\label{lam}
\and Department of Astronomy, Oskar Klein Centre, Stockholm University, 106 91 Stockholm, Sweden\label{oskar}
\and School of Cosmic Physics, Dublin Institute for Advanced Studies, 31 Fitzwilliam Place, Dublin, D02 XF86, Ireland\label{dublin}}

\date{}

 
  \abstract
   {Mid-infrared imaging of exoplanets and disks is now possible with the coronagraphs of the Mid-InfraRed Instrument (MIRI) on the James Webb Space Telescope (JWST).
   This wavelength range unveils new features of young directly imaged systems and allows us to obtain new constraints for characterizing the atmosphere of young giant exoplanets and associated disks.} 
   {These observations aim to characterize the atmosphere of the planet HD\,95086\,b by adding mid-infrared information so that the various hypotheses about its atmospheric parameters values can be unraveled. Improved images of circumstellar disks are provided.}
   {We present the MIRI coronagraphic imaging of the system HD\,95086 obtained with the F1065C, F1140, and F2300C filters
   at central wavelengths of 10.575\,$\mu$m, 11.3\,$\mu$m, and 23\,$\mu$m, respectively.
   We explored the method for subtracting the stellar diffraction pattern in the particular case when bright dust emitting at short separation is present.
   Furthermore, we compared different methods for extracting the photometry of the planet.
   Using the atmospheric models \texttt{Exo-REM} and \texttt{ATMO}, we measured the atmospheric parameters of HD\,95086\,b.}
   {The planet HD\,95086\,b is detected at the two shortest MIRI wavelengths F1065C and F1140C. The contribution from the inner disk of the system is also detected. It is similar to that in the HR\,8799 system.
   The outer colder belt is imaged at 23 $\mu$m.
   Background objects are observed in all filters.
   The mid-infrared photometry provides better constraints on the atmospheric parameters. We evaluate a temperature of 800--1050\,K, consistent with one previous hypothesis that only used near-infrared data.
   The radius measurement of 1.0 -- 1.14 R$_{Jup}$ is better aligned with evolutionary models, but still smaller than predicted.
   These observations allow us to refute the hypothesis of a warm circumplanetary disk.}
   {HD\,95086 is one of the first exoplanetary systems to be revealed at mid-infrared wavelengths. This highlights the interests and challenges of observations at these wavelengths.}
   
\authorrunning{M. Mâlin et al.}
\titlerunning{Unveiling the system HD\,95086 at mid-infrared wavelengths with JWST/MIRI.}

\maketitle
%
\section{Introduction}
Directly imaged planets represent only a small fraction of the population of exoplanets detected so far. 
Nevertheless, they correspond to a distinct population of long-period massive giant planets that cannot be accessed with any other method.
Each detection and characterization of these systems is therefore crucial for understanding planetary formation and evolution in general.
Since the first imaged planets, most observations have been made at near-infrared wavelengths with ground-based facilities. 
However, it is crucial to extend the wavelength coverage to the mid-infrared \citep[which has proved to be difficult from the ground;][]{skaf_beta_2023, wagner_imaging_2021, pathak_high-contrast_2021} to better constrain atmospheric properties of exoplanets  and to provide access to several molecular species \citep{miles_jwst_2023, danielski_atmospheric_2018}.

The young and nearby system HD\,95086 is a benchmark for characterizing the architecture of young systems and exoplanetary atmospheres.
The host star is an A8-type star located at a distance of 86.46\,$\pm$\,0.14 pc \citep{gaia_collaboration_gaia_2023}.
The stellar association to which it belongs is debated, and so is its age.
Initially, it was considered as a member of the Lower Centaurus Crux (LLC) population based on its kinematic properties and position in the color-magnitude diagram by \cite{rizzuto_multidimensional_2011}, who estimated an age of 17$\pm$4 Myr.
The membership of HD 95086 to this association was subsequently revised to the Carina system by \cite{booth_age_2021}, together with a proposed younger age for Carina (13.3$^{+1.1}_{-0.6}$\,Myr).
This age value has recently been revised to 41$^{+5}_{-3}$\,Myr based on the lithium-depletion boundary method \cite{wood_lithium_2023}.
However, \cite{wood_tess_2023} found that the star HD\,95086 is instead a high-probability member of their newly discovered MELANGE-4 association (located near the LCC population in the southern part of Sco-Cen, which is at the western edge of the Carina association).
This association is older than the LCC population and is estimated to be $\sim$\,27\,$\pm$ 3\,Myr.

The system harbors a planetary belt structure, and one giant planet was detected between a warm inner dust belt and a broad, colder outer disk.
HD\,95086\,b has been detected at near-infrared wavelengths with VLT/NaCo \citep{rameau_discovery_2013, rameau_confirmation_2013}.
Its mass was initially estimated to be 4--5 M$_{Jup}$, but the older age of the system (27$\pm$3\,Myr) suggests a higher mass of 7.2 \,$\pm$\,0.7 M$_{Jup}$ \citep{wood_tess_2023}.
The near-infrared spectra obtained with both Gemini/GPI and VLT/SPHERE \citep{galicher_near-infrared_2014, de_rosa_spectroscopic_2016, chauvin_investigating_2018,desgrange_-depth_2022} measured the near-IR spectral energy distribution of the planet, which is well fit by spectral models of dusty and/or young L7--L9 dwarfs. 
\cite{desgrange_-depth_2022} investigated two scenarios using spectra from VLT/SPHERE combined with archival observations from VLT/NaCo and Gemini/GPI.
Either the red color of the planet can be explained by the presence of a circumplanetary disk (CPD), with a range of high-temperature solutions (1400 -- 1600\,K) and significant extinction, or it can be explained by a supersolar metallicity atmosphere but lower temperatures (800 -- 1300\,K), and a small to medium amount of extinction.
\cite{rameau_constraints_2016} used VLT/NaCo data from 2012 and 2013 and GPI astrometry monitoring from 2013 to 2016 to constrain the orbital parameters of the planet. They found a semimajor axis of 62$^{+21}_{-8}$\,au, an eccentricity lower than 0.21, and an inclination of 153$^{+10}_{-14}$ degrees.
\cite{chauvin_investigating_2018} combined VLT/NaCo and VLT/SPHERE astrometric measurements to cover a larger fraction of the orbit
and found a period of about 289$^{+12}_{-177}$ years, a semimajor axis of 52$^{+13}_{-24}$ au (for a measured distance of the system of 83.8\,pc), a relatively low eccentricity of 0.2$^{+0.3}_{-0.2}$,  and an inclination of $141^{+15}_{-13}$ degrees that is compatible with a coplanar orbit with the debris disk plane.
More recently, \cite{desgrange_-depth_2022} found consistent results (semimajor axis of 51--73\,au and an  eccentricity lower than 0.18) by using two independent methods based on Markov chain Monte Carlo (MCMC) with data from 2012--2016, and the K-Stacker algorithm with data from 2016--2019. 

The two planetesimal belts were inferred from spectral energy distribution (SED) studies, first with the Herschel Space Observatory and then by combining Spitzer, WISE, and APEX observations \citep{moor_resolved_2013, su_debris_2015}.
Only the outer belt was resolved using the Atacama Large Millimeter Array (ALMA) in the continuum at 1.3\,$\mu$m.
According to ALMA imaging, the inner and outer edges of the cold belt ($\sim$ 55\,K) are located at
106 $\pm$ 6\,au and 320\,$\pm$\,20 au from the star \citep{su_alma_2017}.
This broad disk has also been tentatively resolved in polarized scattered light with VLT/SPHERE \citep{chauvin_investigating_2018}.
Closer in, the location of the asteroid-like belt ($\sim$ 175\,K) is estimated to lie at 8 $\pm$ 2\,au, but has never been imaged so far.
Farther out, a halo of small dust particles extends to 800\,au \citep{su_debris_2015}.
Another innermost belt has been hypothesized to account for IR excess shortward of 10\,$\mu$m \citep{su_debris_2015}, which might be located within 2\,au and at a temperature of 300\,K. 
This very highly structured pattern led \cite{su_debris_2015} to propose scenarios with up to four planets.
%
The planet HD\,95086\,b likely sculpts the inner edge of the outer ring, but cannot sustain the large cavity observed between 10 and 106\,au alone. At the same time, the planetesimal-belt architecture also suggests the presence of additional planets, which remain to be identified \citep{su_alma_2017, rameau_constraints_2016}.
To identify them, \cite{chauvin_investigating_2018} used the High Accuracy Radial velocity Planet Searcher (HARPS) high-resolution optical spectrograph to search for additional exoplanets with the radial velocity (RV) technique. They were able to exclude the presence of a very massive ($>$10\,M$_{Jup}$) coplanar inner giant planet at less than 1\,au.
Furthermore, \cite{desgrange_-depth_2022} took advantage of the K-Stacker algorithm to push the detection performance by combining  several epochs together and exploiting the fact that the Keplerian motion can allow us to disentangle residual stellar light from a real exoplanet. No robust candidate was found. 
Nonetheless, new constraints on the masses and locations of putative additional exoplanets in the system were derived, and they ruled out any other 5\,M$_{Jup}$ inner planet in the system that would be located at a distance larger than 17\,au at the 50\% confidence level (or a 9\,M$_{Jup}$ inner planet at a distance larger than 10\,au at the 50\% confidence level).

Previous studies of the HD\,95086\,b planet were all performed with near-IR ground-based instruments, but JWST/MIRI opens the window to mid-infrared wavelengths observations.
The MIRI coronagraphs were designed to allow NH$_3$ detection (F1065C and F1140C), to provide independent temperature measurements (e.g., combining F1140C and F1550C), and to image cold circumstellar disks \citep{boccaletti_mid-infrared_2015}.
The results from the Early Release Science (ERS) program \citep[PID 1386]{hinkley_jwst_2022} confirmed their great performances by imaging of the first exoplanet at mid-infrared wavelengths, the giant gaseous exoplanet HIP\,65426\,b \citep{carter_jwst_2023}.
Moreover, MIRI images of HR\,8799 revealed the four giant planets at 10.5 and 11.3 $\mu$m, and the image at 15.5\,$\mu$m presents the first spatially resolved detection of the inner warm debris disk \citep{boccaletti_imaging_2024}.
These very first MIRI observations also highlighted the challenges that arise from contamination by background objects in mid-infrared observations.
The system HR\,8799 is one of the most frequently studied systems and was often considered as a young Solar System analog, with its two debris belts and four giant planets in between \citep{marois_direct_2008, marois_images_2010}. 
The structure of HD\,95086 is similar to that of HR\,8799, in which only one giant planet has been detected.
Therefore, we can expect similar challenges and results for the HD\,95086 system.

We present the first mid-infrared observations obtained with JWST/MIRI of the system HD\,95086\,b as part of the ExoMIRI Guaranteed Time Observations (GTO 1277, PI: P.-O. Lagage).
This source is one of the sources in MIRIco, a EU-US coordinated observing effort with the MIRI coronagraphs between programs 1194, 1277, and 1241.
The goal of this program is to obtain observations at mid-infrared wavelengths to extend the wavelength range from the visible to the mid-infrared to characterize young exoplanetary systems. 
Section \ref{sec:obs_data_red} presents the observation settings and the data reduction. Section \ref{sec:photom} describes the methods for extracting the photometry of HD\,95086\,b, and Sect. \ref{sec:spectral_charact} focuses on characterizing its atmosphere. 
Section \ref{sec:debris_disks} displays the analysis of the debris disks in the system.
Finally, we discuss our results in  Sect. \ref{sec:discusion}, and Sect. \ref{sec:conclusion} summarizes the conclusion and perspective of our study.

\section{Observations and data reduction}
\label{sec:obs_data_red}
\subsection{Program observations}
Observations were conducted using the MIRI 4-Quadrant Phase-Mask (4QPM) coronagraphs along with the F1065C and F1140C filters, as well as the Lyot coronagraph with the F2300C filter \citep{rouan_fourquadrant_2000, boccaletti_mid-infrared_2015}.
We refer to \cite{boccaletti_jwstmiri_2022} for their on-sky performances.
A background observation is included for each filter to mitigate the so-called glowstick, which was identified during the commissioning \citep{boccaletti_jwstmiri_2022}.
In addition, a reference star is observed in the same modes to remove the stellar diffraction residuals left unattenuated by the coronagraphs.
For the F1065C and the F1140C filters, reference star observations are performed with the Small-Grid Dither (SGD) setup \citep{lajoie_small-grid_2016}. 
The reference star is observed nine times, incorporating a subpixel offset to account for the inherent challenge of accurately measuring the center of the coronagraph masks.
This method provides ample diversity to reconstruct an optimized reference coronagraphic image with optimized linear combinations or principal component analysis (PCA) using the various reference frames.
For the F2300C filter, the mask is much larger (an inner working angle of 2.16$''$, in comparison to 0.33$''$ for the 4QPM at F1065C, e.g.),
and therefore, stellar residuals are a smaller problem than the background level.
In these conditions, a single observation of the reference star is sufficient to achieve a satisfactory subtraction of the stellar diffraction pattern.
The reference star was chosen to have an angle close to HD\,95086 ($<$ 1$^\circ$). 
Because nine observations are required, the reference star (chosen to be about three times brighter in the WISE:W3 filter) was observed with fewer groups to reduce the time for each dither position.
All observation parameters are summarized in Tab. \ref{tab:log_obs} and the coordinates RA/DEC of each source and associated background are indicated in Tab. \ref{tab:Coordinates}.
\begin{table*}[h!]
\begin{center}
\caption{Observation parameters of the HD\,95086 system as part of the MIRI GTO program 1277.}
\begin{tabular}{lllllllll}
\hline
\hline
Date and time  & Filter & Object   &  Type     & Obs ID  & N$_{group}$ & N$_{int}$ & N$_{dither}$ & T$_\mathrm{exp}$                          \\
 UT         &   &          &           &         &             &           &        &  per dither (s) \\  
\hline
\hline
16/03/2023 03:30 & F1065C & HD\,95086    & Target on    & obs 1  & 500 & 16  & 1 & 1921.035 \\
16/03/2023 07:31 & F1065C & $-$          & Background   & obs 6  & 500 & 8   & 2 & 1920.796 \\
16/03/2023 08:40 & F1065C & HD\,310459   & Ref on       & obs 7  & 150 & 9   & 9 & 2929.369 \\ 
16/03/2023 13:08 & F1065C & $-$         & Background   & obs 12 & 150 & 9   & 2 & 650.971 \\
\hline
16/03/2023 04:22  & F1140C & HD\,95086    & Target on    & obs 2  & 500 & 16  & 1 & 1921.035 \\
16/03/2023 06:48  & F1140C & $-$          & Background   & obs 5  & 500 &  8  & 2 & 1920.796 \\
16/03/2023 09:54  & F1140C & HD\,310459   & Ref on       & obs 8  & 150 & 9   & 9 & 2929.369 \\
16/03/2023 12:47 & F1140C & $-$          & Background   & obs 11 & 150 & 9   & 2 & 650.971 \\
\hline
16/03/2023 05:18  & F2300C & HD\,95086    & Target on    & obs 3 & 30 & 200  & 1 & 2008.476 \\
16/03/2023 06:02 & F2300C & $-$          & Background   & obs 4 & 30 & 100  & 2 & 2008.152 \\
16/03/2023 11:13 & F2300C & HD\,310459   & Ref on       & obs 9 & 30 & 200   & 1 & 2008.476 \\
16/03/2023 12:01 & F2300C & $-$         & Background   & obs 10 & 30 & 100   & 2 & 2008.152\\
\hline
\end{tabular}
\tablefoot{Date and time represent the starting time of the observation on the target, followed by the order of the execution of the sequence, the filter, the name of the object, the type, and the ID of each observation.
The last parameters represent the instrument observation settings: The 
number of groups, the number of integrations, the number of dither positions, and the total exposure time per dither.}
\label{tab:log_obs}
\end{center}
\end{table*}

\begin{table}[]
\caption{Coordinates of each observation. Coordinates are indicated in epoch 2000. }
    \centering
    \begin{tabular}{c|cc}
        \hline
        \hline
        Source & RA & DEC\\
        \hline
        HD\,95086 &  10 57 03.0200 & -68 40 02.40\\
        HD\,95086 Background & 10 57 06.7600 & -68 43 20.80\\
        HD\,310459 & 11 14 09.6246 & -68 34 11.57\\
        HD\,310459 Background & 11 15 38.2600 & -68 36 00.20 \\
        \hline
    \end{tabular}
    \tablefoot{Right ascension (RA) is expressed in hours, minutes, and seconds, and the declination (DEC) is listed in degrees, minutes, and seconds.}
    \label{tab:Coordinates}
\end{table}


\subsection{Data reduction}
\label{sec:data_reduction}
\begin{figure*}[h]
    \centering
    \includegraphics[width=18cm]{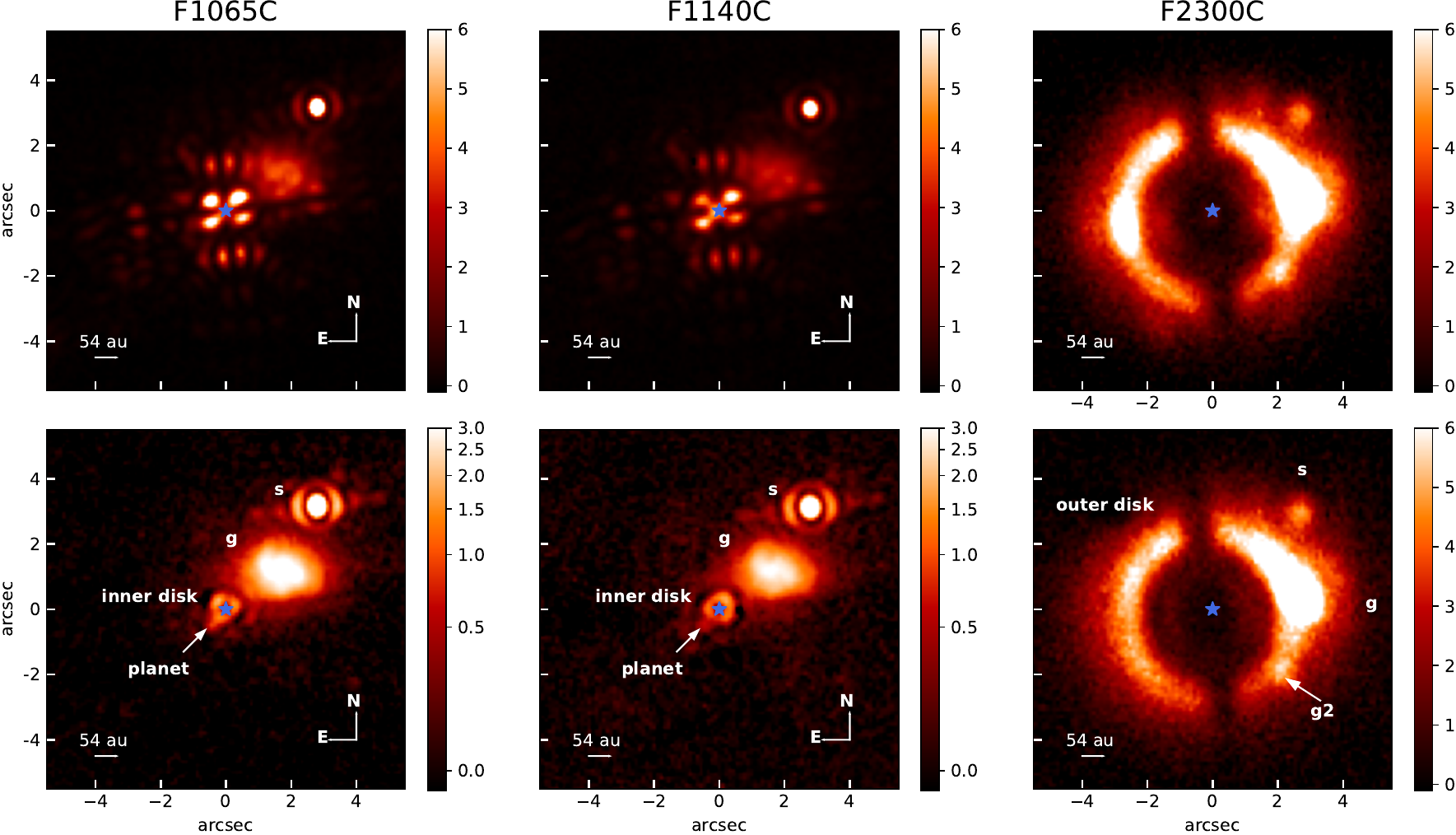}
    \caption{Coronagraphic images of HD\,95086 in each filter (F1065, F1140, and F2300) are shown in the top panel, followed by the same images after the subtraction of an optimized-reference star in the lower panel.
    The coronagraphic center is illustrated with the small star in blue. The inner disk, the planet, the background galaxies g and g2, and the background star s are labeled.
    For clarity, the  color scale is enhanced for the bottom pannel at F1065C and F1140C, using an asinh scale. Units are in DN/s.}
    \label{fig:image_sub}
\end{figure*}

The data were retrieved from the Mikulski Archive for Space Telescopes (MAST)\footnote{MAST : \href{https://mast.stsci.edu/portal/Mashup/Clients/Mast/Portal.html}{mast.stsci.edu}}.
Starting from the uncalibrated data (\texttt{$\_$uncal}), we ran stage 1 of the \texttt{JWST} pipeline \footnote{\texttt{JWST} pipeline : \href{https://jwst-pipeline.readthedocs.io/en/latest/jwst/package_index.html}{jwst-pipeline.readthedocs.io}, version : 1.12.5, CRDS = 1140}, which applies essential detector-level corrections to all exposure types to obtain a corrected count-rate image.
As noted by \cite{carter_jwst_2023}, some pixels are erroneously flagged as containing a jump because the default threshold value for identifying jumps is too low.
We had 500 groups per integration, and the impact of the jump threshold was therefore weaker than in ERS data. We note that it affects the data reduction very little. 
Furthermore, to assess the impact of the data reduction on the planet photometry, we led our full analysis with various reductions, including \texttt{spaceklip}\footnote{\texttt{spaceklip} : \href{https://spaceklip.readthedocs.io/en/latest/}{spaceklip.readthedocs.io}} \citep{kammerer_performance_2022}, which calls the \texttt{JWST} pipeline with customize bad-pixel correction, and their best parameters were evaluated with ERS data from \cite{carter_jwst_2023}.
We verified that the first integration did not deviate from other integrations, as identified in observations of HIP\,65426\,b.
While the first integration presents a slightly higher flux level, the median flux does not deviate from the average median flux in all integrations (lower than 1\%).
Compared to the ERS observations, GTO observation programs use fewer integrations per observation, so that removing the first integration would significantly decrease the S/N.
At stage 2 of the \texttt{JWST} pipeline, we skipped the \texttt{flat$\_$field} step to avoid increasing the noise level near the transitions of the 4QPMs. 
The pixels that are close to these transitions are attenuated by the coronagraph, but this is the result of a diffraction effect and not just a transmission effect.
The field-dependent attenuation of the coronagraph was considered when we extracted the photometry of point sources (see Sect. \ref{sec:extract_photom}).
In addition, we ran stage 2 both with and without the \texttt{photom} step because one of our goals is to evaluate and compare the photometric calibration using various methods.
The mean of the two background exposures was then subtracted from each science image.
We applied a $\sigma$-clipping function to all pixels whose value was greater than 3$\sigma$ compared to the median of a region with a radius of the two nearest neighbors, and we replaced it by this median value. 
The NaN values were also corrected in the same way.
Finally, the images were rotated using the position angle 
(angle of $\sim$ 12$^\circ$) to align north with the top of the image.
The final images are shown in the top panel of Fig. \ref{fig:image_sub}.
Stellar diffraction dominates the images in the F1065C, F1140C, and F2300C  filters.
In the F2300C filter, the stellar diffraction is still visible, but the predominant contribution appears to be from the outer disk belt (see Sect. \ref{sec:outer_disk}).
Two background objects are clearly identified in the NW of the images and are discussed further in Sect. \ref{sec:images_description}.

\subsection{Reference star subtraction}
\label{sec:stellar_subtraction}
To further suppress
the diffraction pattern of the star in the coronagraphic images, we subtracted an optimized reference star image.
The various methods we applied are presented in Appendix \ref{app:other_red}.
We first applied reference star differential imaging (RDI) methods such as PCA \citep[similar to KLIP algorithm,][]{soummer_detection_2012} and linear optimization of the reference star observations. 
As a first test, we applied a linear combination of the nine reference star images using the Nelder-Mead simplex with the Python function \texttt{scipy.optimize.fmin} \citep{nelder_simplex_1965}.
The images obtained with these algorithms produce strong negative residuals with a similar shape as the diffraction pattern of the star (Fig. \ref{fig:HD95_opt_lin}). 
The stellar diffraction was still oversubtracted \citep{pueyo_detection_2016}, 
even when we masked the background objects to ensure that they did not impact the minimization process.
As a result, a faint source is discernible at the expected location of the planet in the F1065C filter alone.
In addition, an inner emission appears, which can be identified as originating from the inner disk that was previously inferred from the SED of the system \citep{su_debris_2015}.
Second, we also tested the PCA method
(Fig. \ref{fig:HD95_PCA}).
The truncation of the principal components has a direct impact on the structures that can be seen in the resulting image.
The inner emission also appears in the F1065C and F1140C filters. 

In conclusion, conventional PCA and linear combination methods do not effectively suppress the stellar diffraction and do not easily allow us to identify the planet HD\,95086\,b.
The RDI subtraction based on the 9-SGD method seems to overestimate the stellar diffraction,
as we can note negative residuals similar to the diffraction pattern after subtraction.
We suspect that because of the inner disk in the HD\,95086 system, the intensity of which is significant at these wavelengths, the PCA and linear combination reconstruct a reference star frame with a higher flux level than the observed star, even when the background sources seen in the upper right part of the field of view are masked.
As a result, the planet is only faintly detected, and the flux estimate is likely to be underestimated.
In addition, we used different observations of available reference stars: ERS 1386 \citep{carter_jwst_2023}, GTO 1194 \citep{boccaletti_imaging_2024}, and commissioning data \citep{boccaletti_jwstmiri_2022}. 
We note that using these observations to expand the reference library did not provide a more effective stellar subtraction (see appendix Fig. \ref{fig:HD95_lib_opt_lin}).\\

Therefore, we implemented a simpler subtraction method to minimize oversubtraction and to increase the signal-to-noise ratio at the planet location. 
For the nine reference observations, we measured a rescaling factor with the target star image that minimized the residuals within a specified masked area with the least-squares method.
We assessed the impact of the masked area by experimenting with various mask shapes. 
A better subtraction was achieved using a mask in the form of a 1$''$ to 2.5$''$ ring, and this also ensured that background objects were masked.
The mask is shown in Fig. \ref{fig:mask_region_min_F1065}.
This region was chosen to exclude the contributions from the planet, the inner disk, and background objects. 
Minimization was therefore performed exclusively within the second stellar diffraction ring.
By employing this method, we mitigated oversubtraction and ensured an effective subtraction of the stellar diffraction pattern. 
Consequently, the planet became visible in images from the F1065C and F1140C filters.
The final result is presented in Fig. \ref{fig:image_sub} in the lower panel.
In summary, we used only the first reference image at F1065C and the fourth reference image at F1140C and applied rescaling factors of 0.304 and 0.312, respectively. 
For the F2300C filter, for which only one reference star is available, the best subtraction was achieved using a rescaling factor of 0.57. 
In this latter case, we chose to minimize the residuals inside a region ranging from 3$''$ to 8$''$ from the center.

\subsection{Image descriptions and background objects}
\label{sec:images_description}
Even though the planet is faint, it is well detected in F1065C and in F1140C, as indicated in the lower panel of Fig. \ref{fig:image_sub}.
In addition, we observe an inner emission at these same wavelengths that is due to the inner disk, similar to the one observed at F1550C in the HR\,8799 system \citep{boccaletti_imaging_2024}.
In the F2300C filter, the image is mostly dominated by the emission of the outer disk,
but the reference star subtraction still allowed us to remove some faint diffraction structures. 
The outer disk contributes most at the longest wavelengths.
However, the Lyot mask has a radius of 2.16$''$ and hence occults a substantial area in the center of the image.

In the F1065C and F1140C filters, two background objects are clearly visible in the field of view: a point source (labeled s), and an extended source (labeled g).
In the F2300C filter, these two same background objects are also clearly identified, as is a third faint object (labeled g2).
The position of each background source and aperture photometry for the background star were measured and are indicated in Tab. \ref{tab:bkg_source}.
\begin{table}[h]
    \caption{Position of the photocenter with respect to the center of the coronagraph in each filter.}
    \centering
    \begin{tabular}{c|c|c c c}
        \hline
        \hline
        Filter & Source & $\Delta$RA ($''$) & $\Delta$Dec ($''$) & Flux (W/m$^2$/$\mu$m)\\
        \hline
         F1065C & s & -2.695 & 3.05 & 3.33$\times$10$^{-18}$\\
                & g & -1.485 & 1.1& \\
        \hline
         F1140C & s & -2.695 & 3.05 & 2.46$\times$10$^{-18}$\\
                & g & -1.485 & 1.1& \\
        \hline
         F2300C & s  & -2.695 & 2.92 & 2.22$\times$10$^{-18}$\\
                & g  & -2.64 & 0.33&\\
                & g2 & -2.035 & -1.815&\\
        \hline
        \end{tabular}
        \tablefoot{Flux values correspond to aperture photometry centered on the measured position of the source.}
    \label{tab:bkg_source}
\end{table}

The first object (designated s in Fig. \ref{fig:image_sub}) is located at the same position in all three filters, and it is compatible with VLT/SPHERE astrometric measurements \citep[specifically source cc-8 in Fig. 8 of][]{chauvin_investigating_2018}. Its flux at F1065C and F1140C would be consistent with an M star, although the flux at F2300C is too high even when the outer disk contribution is subtracted out 
(using the disk flux measured at the same distance, but on the opposite side of the mask). If it were an unresolved background galaxy, a detailed study would be necessary to confirm whether this is consistent with the photometry.

The second object is identified as a background galaxy (labeled g in Fig. \ref{fig:image_sub}), consistent with the observation from ALMA at 1.3\,mm \citep{su_alma_2017}. 
MIRI even allows us to resolve its spiral structure.
The contamination by background galaxies in this system was already expected from \cite{su_debris_2015} because the integrated submillimeter flux presents an excess emission in Herschel/SPIRE bands and APEX 870 $\mu$m compared to the disk SED model. 
\cite{su_alma_2017} concluded that this likely is a dusty star-forming galaxy at $z = 2$.
Based on its brightness in the dust continuum and on the nondetection of the CO emission, we can definitely rule out a structure that is physically associated with the system (e.g., a dust clump in the disk).
The nature of this object has also been confirmed based on the absence of proper motions \citep{zapata_submillimeter_2018} and with the lack of a concentration of CO at the location of the bright source and the clear difference in the spectral index between the source and the disk \citep{booth_deep_2019}.
This background galaxy has never been detected in any previous near-IR observations \citep[e.g. ][]{rameau_confirmation_2013, chauvin_investigating_2018, desgrange_-depth_2022}.
In the F2300C filter, the large opaque Lyot mask occults the central region of the galaxy, which explains the apparent mismatch of the centroid position with respect to F1065C and F1140C.

The third faint point source detected at F2300C was also identified in the ALMA observations at 1.3 mm (denoted g2 in Fig. \ref{fig:image_sub}).
It is likely a background galaxy, similar to other faint sources observed in the long-baseline map \citep{su_alma_2017}.

Finally, another faint point source next to planet b seems to appear in the F1140C filter
(see the zoom-in in Fig \ref{fig:model_new_source}, in the SE diagonal from the center of the image).
This is discussed in Sect. \ref{sec:add_planets}.

\section{Photometry of the planet}
\label{sec:photom}
\subsection{Extraction of the photometry}
\label{sec:extract_photom}
We used the website \url{whereistheplanet}\footnote{\href{http://whereistheplanet.com/}{whereistheplanet.com}} \citep{wang_whereistheplanet_2021} to provide an estimate of the planet position at the observation date that relied on previous observations.
The photometry of the planet can be estimated using both aperture photometry and a modeling of the point spread function (PSF) of the planet. 
For the MIRI coronagraphic observations of HR\,8799, an aperture size of $1.5\,\lambda/D$ was found to be the best fit for integrating the planetary signal \citep{boccaletti_imaging_2024}.
However, for HD\,95086, this aperture size resulted in a significant overestimate of the planet flux, even with a smaller aperture (down to $1\,\lambda$/D). 
Therefore, we conclude that the aperture photometry is unreliable here because the contribution of the inner disk overlaps the planet PSF.\\

\begin{figure*}[h]
    \centering
    \includegraphics[width=18cm]{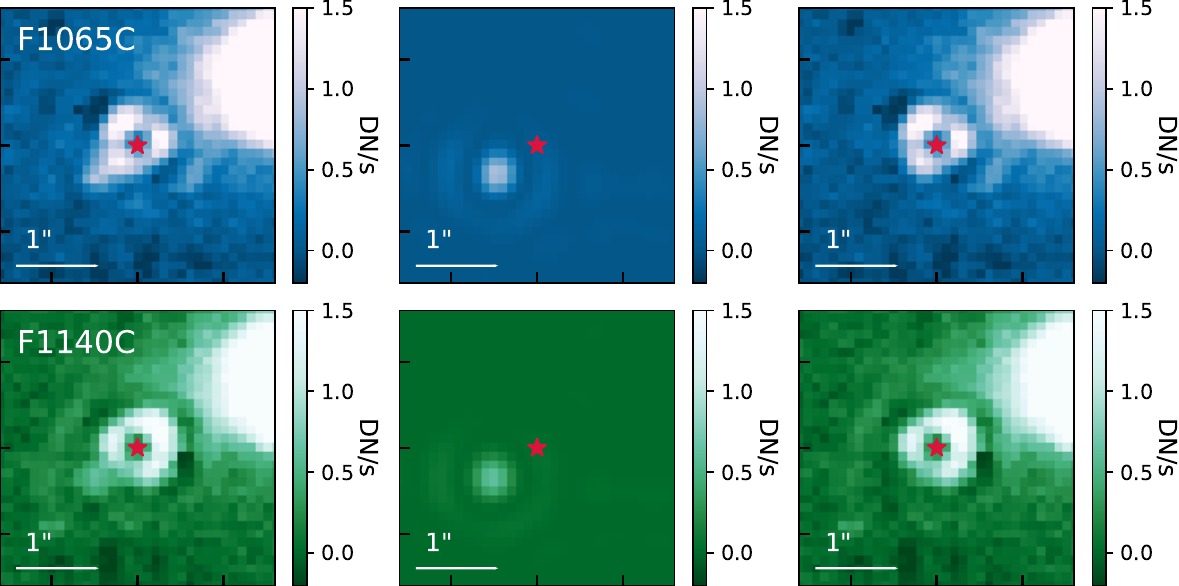}
    \caption{Zoom-in of the planet in the final image in the detector frame coordinates. 
    Left : Data after subtraction of the reference star. 
    Middle : Model of the generated planet PSF. 
    Right : Residuals of the data minus the model. 
    The center of the coronagraph masks is represented by the red star. The top panel corresponds to F1065C, and the bottom panel corresponds to F1140C.}
    \label{fig:planet_model}
\end{figure*}

Consequently, it became necessary to model the planet PSF to accurately obtain its photometry.
We used \texttt{WebbPSF}\footnote{\texttt{WebbPSF} : \href{https://webbpsf.readthedocs.io/en/latest/}{webbpsf.readthedocs.io}} \citep{perrin_updated_2014} to compute synthetic PSF images,
taking into account the appropriate filter and mask configurations for MIRI coronagraphs.
A PSF was simulated for each filter F1065C and F1140C with the corresponding mask FQPM1065 and FQPM1140, and the pupil mask MASKFQPM. 
The planet position with respect to the 4QPM axis was taken into account by specifying its sky coordinates, the V3PA angle, which defines the observatory orientation, and the inclination of the 4QPM with respect to the detector (4.835$^{\circ}$).
The simulated PSF was then used to inject a negative simulated planet into the data \citep{lagrange_probable_2009}.
We derived the position and the flux of the planet by minimizing the residuals with the Nelder-Mead \citep{gao_implementing_2012} algorithm (Python function \texttt{scipy.optimize.minimize}) within an aperture of $1.2\,\lambda$/D at the planet position.
A mask with a radius of 5.5 pixels ($\sim$ 0.6$''$, located at the center of the coronagraphs) was used to ensure that the minimization was not  biased by the disk.
In the minimization process, we bounded the parameters to a maximum variation of $\pm$\,1 pixel and to a flux level between 0 and 5\,DN/s.
For the priors, we took the expected position of the planet and an arbitrary flux of 1\,DN/s.
The best-fit planet PSF models are displayed in the middle panels in Fig. \ref{fig:planet_model}.
The residuals after subtracting the model from the data are shown in the right panels in Fig. \ref{fig:planet_model}. 
This figure confirms that the planet is well subtracted, avoiding disk contamination, and only the inner disk contribution remains in the image (see Sect. \ref{sec:inner_disk}).
The flux for the planet was then measured on the model PSF.

The following step was made to evaluate the attenuation of the coronagraph at the position of the planet. 
For this purpose, two PSFs were simulated with \texttt{WebbPSF} (using the same configuration as for modeling the planet PSF). 
The first PSF was at the planet position, and the second farther away with an offset of 10" that was not affected by the coronagraph attenuation.
The fluxes of both PSFs were measured in the same aperture, the ratio of which provides a measure for the intrinsic coronagraph attenuation at the planet position.
The attenuation is 0.63 in F1065C and 0.62 in F1140C, respectively. 
The flux values measured in the images were then divided by the attenuation.
The final values of the flux emitted by the planet are listed in Tab. \ref{tab:exctrated_flux_values}.
Similar values are obtained with the diffraction model from \citet{boccaletti_mid-infrared_2015}, which was calibrated with commissioning data.
\begin{table}[H]
    \centering
    \caption{Extracted photometry for HD\,95086\,b with and without the photometric calibration.}
    \begin{tabular}{c|c c}
        \hline
        \hline
        Extracted values & F1065C & F1140C \\
        \hline
         without photometric calibration in DN/s & 41.4 & 32.5\\
         with photometric calibration in MJy/sr & 127.7 & 91.8 \\
        \hline
        \end{tabular}
    \label{tab:exctrated_flux_values}
\end{table}

\subsection{Normalization to flux units}
\label{sec:normalisation}

Similar to \cite{boccaletti_imaging_2024}, we applied a contrast-based normalization to obtain the calibration to physical flux units for the dataset without the \texttt{photom} stage (therefore still in count rate).
The goal was to compare our results with the photometric calibration that was recently included in the pipeline (reference file crds above the version 1140). 
Coronagraphic observations with ground-based instruments usually rely on the observations of the stellar PSF off-axis in order to measure the contrast of the planet with respect to its host star.
The stability of the space telescopes means that the PSF is more stable and could be estimated without an additional observation of the star off-axis.
As a consequence, we were able to either use previous observations to estimate the stellar off-axis flux or use a direct conversion from the DN/s to physical units
(Tab. \ref{tab:exctrated_flux_values}). 
We used the commissioning data (reduced with the method described in Sect. \ref{sec:data_reduction}) to derive a normalization factor between the star used in commissioning (HD\,15816 at F1065C and F1140C, and HD\,163113 at F2300C) and HD\,95086.
This allowed us to estimate an off-axis PSF for the target star, which was used to measure the contrast between the star and the planet.
The three methods based on the contrast measurement are listed below.\\

1/ The first method was to measure the flux ratio of the target acquisition (TA) images between the target HD\,95086 and the commissioning star and to apply this ratio to the commissioning off-axis PSF to obtain an estimate of an off-axis PSF for HD\,95086.
We note that the values obtained using this method might be biased due to the use of TA, which involves a neutral density filter \citep[as discussed in][]{boccaletti_imaging_2024}.

2/ The second way to obtain a normalization is to measure the ratio of the coronagraphic images (target and commissioning) in the same filter (F1065C, F1140C, or F2300C).
In coronagraphic images, the flux is spread over the field of view, and we therefore needed to integrate the flux in a large aperture of 15\,$\times\lambda$/D. 
However, for the HD\,95086 system, this approach might yield inaccurate results because of the background objects, which might result in an overestimation of the stellar flux.
Furthermore, even when we masked the background objects in the commissioning on-axis data and in the HD\,95086 data, the scaling factor was still not reliable. 
Almost one quarter of the image is masked when the background galaxy is masked. Therefore, from one dither position to the next of the 9-SGD sequence, we do not measure the same flux.

3/ Instead of using the commissioning data, as in method 2, we can take advantage of \texttt{WebbPSF} to generate synthetic PSF and coronagraphic images. For each filter, we measured the flux factor that minimized the residuals between the synthetic and actual coronagraphic images while masking the background objects.
The position of the simulated star was assumed to be perfectly centered on the coronagraphic mask.
This synthetic coronagraphic image was then used in the minimization routine, taking into account two free parameters for the image position on the detector. 
We also took the pupil shear of 2.5\% into account \citep[measured with commisioning data,][]{boccaletti_jwstmiri_2022}.
Then, we rescaled the PSF according to this factor in order to normalize the PSF, and this value was used to measure the contrast with the planet.\\

\begin{figure}[h!]
    \centering
    \includegraphics[width=8.8cm]{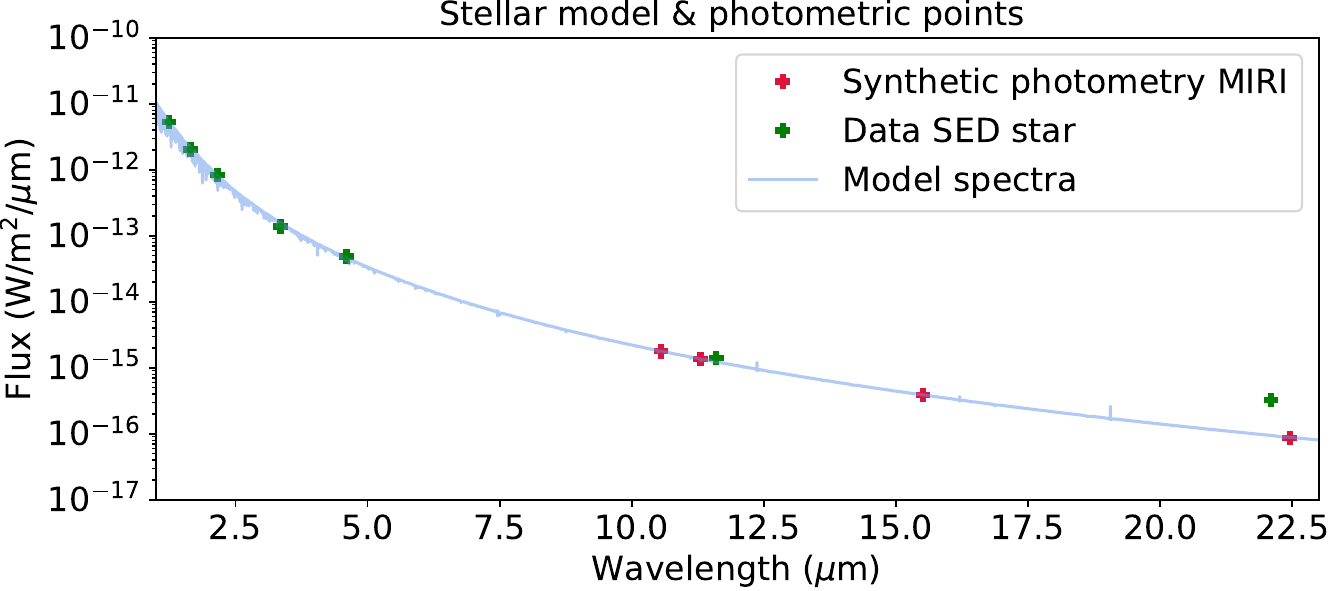}
    \caption{BT-Nextgen model rescaled to the flux level of 
     HD\,95086 (blue), measured photometric values (green), and rescaled synthetic photometry in MIRI filters (red).}
    \label{fig:star_model}
\end{figure}

These three methods enabled us to obtain an estimate of the off-axis PSF in order to measure the contrast with respect to the planet-extracted flux in DN/s. 
We considered a BT-Nextgen stellar model at 7600\,K using VOSA\footnote{VOSA : \href{http://svo2.cab.inta-csic.es/theory/vosa/}{svo2.cab.inta-csic.es/theory/vosa}.} \citep[VO Sed Analyzer,][]{bayo_vosa_2008}.
Further, we assumed the following parameters : 1.5 R$_{\sun}$ for the radius, and 86.46\,pc for the distance \citep{gaia_collaboration_gaia_2023}.
We fit this model on the measured photometric data points of HD\,95086 in the 2MASS filters J, H, K$_s$ and in the WISE filters W1 and W2 
(retrieved from VizieR\footnote{\label{ftnote:vizier}VizieR : \href{https://vizier.cds.unistra.fr/viz-bin/VizieR}{vizier.cds.unistra.fr}.}) to derive a coefficient that minimizes the $\chi^2$ between the stellar model and the stellar photometry. 
Photometric values above 10\,$\mu$m were not taken into account to fit the stellar model, because they might be impacted by the debris disk.
We obtained a scaling factor of 0.99. 
The corresponding stellar photometry in the MIRI filters was also retrieved from VOSA and was rescaled accordingly. 
The stellar tabulated photometric values, the fit model, and the MIRI synthetic values are shown in Fig. \ref{fig:star_model}.\\

The fourth method provides the more straightforward way, using the constant calibration from DN/s to MJy/sr, and then converting into W/m$^2$/$\mu$m, as described in Sect. \ref{sec:extract_photom}.
The photometric values obtained for the planet with each method are provided in Tab. \ref{tab:flux_values}.
The \texttt{JWST} pipeline (with versions above 1.2.5 and a crds context above 1140) gives photometry values in method 4 that agree with those obtained with the contrast-measurement methods 1 and 3. 
We pursued the analysis with the mean and standard deviation of the three reliable methods.
As stated in \cite{boccaletti_imaging_2024}, we identified that one of the primary source of uncertainty in the planet photometry arises from the normalization of the PSF from the coronagraphic image.
We assessed the impact of the inner disk contamination onto the planet flux by measuring the flux in an annulus.
at the planet separation, masking the planet and the galaxy.
Both uncertainties are taken into account in the final uncertainty provided in Tab. \ref{tab:flux_values} : the uncertainties from the flux normalization are 13\% in the F1065C filter and 2\% in the F1140C filter, and those due to the inner disk contamination are 14\% and 12\% in the F1065C and F1140C filters, respectively.
\begin{table}[h!]
    \centering
    \caption{Flux values measured for the planet in W/m$^2$/$\mu$m.}
    \begin{tabular}{c|c c}
        \hline
        \hline
        Method & Flux F1065C & Flux F1140C \\
        \hline
         1 & 7.03 $\times$ 10$^{-19}$ & 6.33 $\times$ 10$^{-19}$\\
         2$^{*}$ & 5.05 $\times$ 10$^{-19}$ & 3.01 $\times$ 10$^{-19}$\\
         3 & 9.16 $\times$ 10$^{-19}$ & 6.40 $\times$ 10$^{-19}$\\
         4 & 9.73 $\times$ 10$^{-19}$& 6.13 $\times$ 10$^{-19}$\\
        \hline
        \end{tabular}
        \tablefoot{$^{*}$Method 2 is unreliable for this system.}
        \label{tab:flux_values}
\end{table}

The final photometry and astrometry of the planet are summarized in Tab \ref{tab:pos_flux}.
The astrometric values were obtained with the negative simulated planet process, which fits the flux and positions of the simulated planet image in the coronagraphic subtracted image.
These values agree with those in previous studies and with the estimate of \url{whereistheplanet}.
The JWST/MIRI astrometric precision obtained with TA observations is $\sim$ 1 mas \citep{rigby_science_2023}.
We assumed that the center of the coronagraphic masks corresponds to the values measured with the commissioning values and provided in the CRDS file.
Therefore, we considered that the stellar position is known to a precision better than $\sim$ 10\,mas \citep{boccaletti_imaging_2024}.
\begin{table}[h!]
    \centering
    \caption{Measured astrometry in arcseconds and photometry for planet HD\,95086\,b.}
    \begin{tabular}{c|c c c}
    \hline
    \hline
        Filters & $\Delta$RA (")  & $\Delta$DEC (") & Flux (W/m$^2$/$\mu$m)\\
    \hline
        F1065C & 0.40 & -0.48 & (8.6 $\pm$ 1.7) $\times$ 10$^{-19}$\\
        F1140C & 0.40 & -0.48 & (6.3 $\pm$ 0.8) $\times$ 10$^{-19}$\\
    \hline
    \end{tabular}
    \tablefoot{The uncertainty on the relative astrometry is $\pm$ 0.01.}
    \label{tab:pos_flux}
\end{table}

Using the various data reductions (with the first pipeline stages processed with \texttt{spaceklip}, different jump threshold, bad pixel processing, etc.; see Sect. \ref{sec:data_reduction}), we find that the variation in the extracted flux for the planet is always within the error bars indicated in Tab \ref{tab:pos_flux}. This means that it has no impact on the atmospheric characterization of the planet.

\section{Spectral characterization of HD\,95086\,b}
\label{sec:spectral_charact}
\subsection{Atmospheric grids}
We used two distinct model grids that are common for characterizing
young directly imaged giant planets \texttt{Exo-REM} and \texttt{ATMO}. 
Their atmospheric parameters are listed in Tab. \ref{tab:param_models_atm}.
\cite{petrus_jwst_2024} recently showed that these models performed well in a fit of the JWST near- and mid-infrared data of VHS\,1256\,b.\\

\noindent{\texttt{Exo-REM}} is a one-dimensional self-consistent radiative-convective equilibrium model.
Initially developed to simulate the atmospheres of young giant planets \citep{baudino_interpreting_2015}, it was later adapted to investigate the L–T transition \citep{charnay_self-consistent_2018} and has more recently been modified for studying irradiated planets \citep{blain_1d_2021}.
The radiative-convective equilibrium is solved by assuming that the net flux (radiative and convective) is conserved. 
The flux conservation is resolved iteratively using a linear inversion method on the pressure grid (64 pressure levels) 
The input parameters are the effective temperature of the planet, gravity, and the elementary abundances of molecules. 
The model includes disequilibrium chemistry \citep{zahnle_methane_2014}.
The cloud model \citep[detailed in][]{charnay_self-consistent_2018} takes into account the microphysics, and the cloud distribution is calculated by taking into account sedimentation and vertical mixing with realistic K$_{zz}$ mixture coefficient profiles based on the mixing length theory. 
The Rayleigh diffusion of H$_2$, He, and H$_2$O, as well as absorption and diffusion by clouds, were taken into account. 
Sources of opacity include collision-induced absorption (CIA) H$_2$--H$_2$, H$_2$--He, H$_2$O--H$_2$O, and H$_2$O-N$_2$, the ro-vibrational bands of molecules (H$_2$O, CH$_4$, CO, CO$_2$, NH$_3$, PH$_3$, TiO, VO, H$_2$S, HCN, and FeH), and the resonant lines of Na and K. 
The molecular line lists used in \texttt{Exo-REM} can be found in \citet{blain_1d_2021}.
In particular, for young planets at long periods, the model incorporates iron and silicate clouds (forsterite). 
The particle radius for these clouds was computed based on a simple microphysics approach within the cloud scheme.\\

\noindent{\texttt{ATMO}}\footnote{\texttt{ATMO} grids are publicly available on the website : \href{https://www.erc-atmo.eu/?page_id=322}{erc-atmo.eu}.} is a one-dimensional atmospheric model \citep{tremblin_cloudless_2017}. 
It considers that clouds are not necessary to reproduce the spectra of brown dwarfs, except for the band of silicates at 10 $\mu$m. 
The phenomenon of fingering convection is proposed as an alternative to clouds \citep{tremblin_cloudless_2016, tremblin_thermo-compositional_2019}.
\texttt{ATMO} assumes that diabatic convection processes induce a disequilibrium chemistry of CO/CH$_4$ and N$_2$/NH$_3$ that can reduce the temperature gradient in atmospheres.
This phenomenon reproduces the reddening of the spectra, which is explained by the presence of clouds in most models, such as \texttt{Exo-REM}.
The model involves an effective adiabatic index $\gamma$ (between 1.01 and 1.05) that controls the change in the temperature gradient. 
The atmospheric pressure levels varied from 2 to 2000 bar
at $\mathrm{log} (g)$ = 5.0 and were scaled for other surface gravities according to $ \times 10^{\mathrm{log} (g) - 5}$.
Disequilibrium chemistry was used with K$_{zz}$=10$^{5}$ cm$^2$/s to $\mathrm{log} (g)$=5.0 and was scaled for other gravity values according to $\times 10^{2(5-\mathrm{log} (g))}$.
The chemistry took 277 species into account, and the disequilibrium chemistry was reproduced according to \citet{venot_chemical_2012}.
Species were included in the hypothesis of local condensation, that is, in the absence of precipitation.
Sources of opacity include H$_2$--H$_2$, H$_2$--He, H$_2$O, CO$_2$, CO, CH$_4$, NH$_3$, Na, K, Li, Rb, Cs, TiO, VO, FeH, PH$_3$, H$_2$S, HCN, C$_2$H$_2$, SO$_2$, Fe, and H-, as well as the Rayleigh diffusion opacities for H$_2$, He, CO, N$_2$, CH$_4$, NH$_3$, H$_2$O, CO$_2$, H$_2$S, et SO$_2$.
We used the \texttt{ATMO} model grid for low-gravity brown dwarf atmospheres presented in \cite{petrus_x-shyne_2023}.
\begin{table}[H]
    \centering
    \caption{Parameters of the atmospheric grids.}
    \begin{tabular}{c|c c}
        \hline
        \hline
        Parameters & \texttt{Exo-REM} &  \texttt{ATMO}\\
        \hline
        \hline
        Temperature (K)  & 400 -- 2000  & 800 -- 3000 \\
                         & Step : 50    & Step : 100\\
        \hline                 
        $\mathrm{Log} (g)$ &  3.0 -- 5.0  & 2.5 -- 5.5 \\
                         & Step : 0.5  & Step : 0.5\\ 
        \hline
        C/O              &  0.1 -- 0.8  &  0.3, 0.55, 0.7\\
                         & Step  : 0.05 & \\
        \hline
        Metallicity [M/H]  & -0.5 -- 1 & -0.6 -- 0.6\\
                         & Step 0.5 & Step : 0.3\\
       \hline
        $\gamma$         &    --      & 1.01, 1.03, 1.05\\
        \hline
        \end{tabular}
    \label{tab:param_models_atm}
\end{table}

\subsection{Near-IR data}
\begin{figure}[h]
    \centering
    \includegraphics[width=9cm]{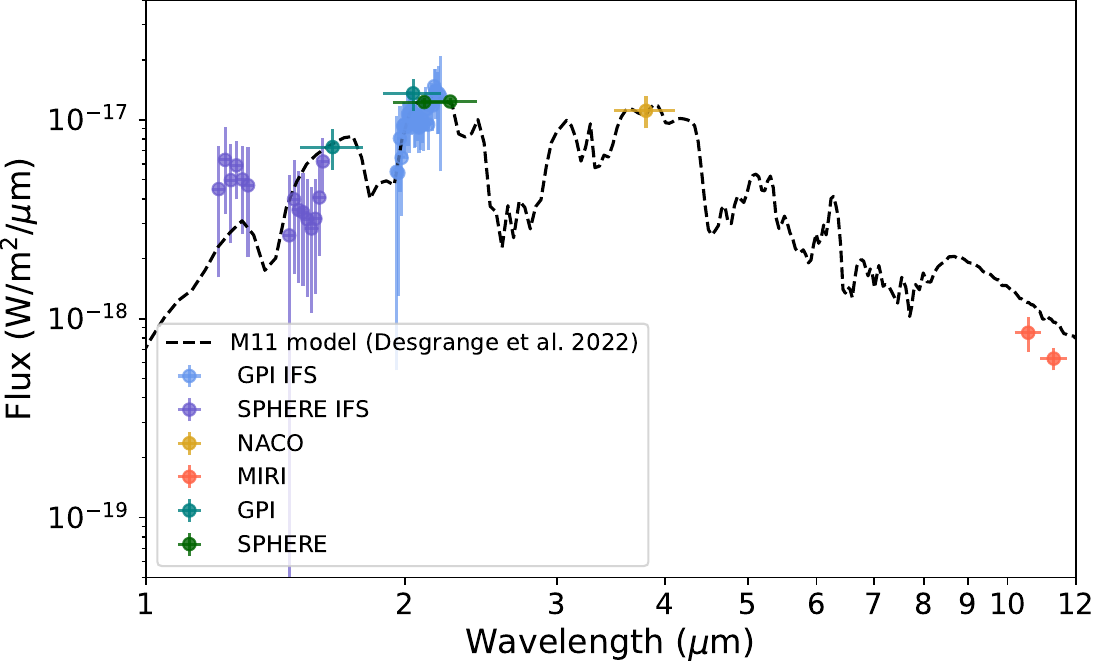}
    \caption{Near-IR photometric and spectroscopic points for HD\,95086\,b. 
    The model M11 \citep[in black,][]{madhusudhan_model_2011}
    corresponds to T = 800\,K, $\mathrm{log} (g)$ = 4.0, R = 1.6 R$_{Jup}$, and [M/H] = 1 without any extinction. It is identified as one of the best-fit models in \cite{desgrange_-depth_2022}.}
    \label{fig:nIR_data}
\end{figure}

In addition to the MIRI photometry,
supplementary archival near-IR photometric data points are available in the literature for the planet HD\,95086\,b:
NaCo in band L' \citep{rameau_discovery_2013, rameau_confirmation_2013}, GPI in  bands H and K \citep{galicher_near-infrared_2014}, and low-resolution spectra \citep{de_rosa_spectroscopic_2016},
SPHERE data in band K1, K2, as well as the low-resolution spectra from SPHERE/IFS \citep{desgrange_-depth_2022}.
The contrast values obtained in these studies were converted into physical flux units using the theoretical stellar model of HD\,95086, following the method outlined in Sect. \ref{sec:normalisation}.
All of these data are displayed in Fig. \ref{fig:nIR_data}, together with the MIRI photometry (Tab. \ref{tab:flux_values}).

\subsection{Forward modeling}
\label{sec:forward_modeling}
To determine the atmospheric parameters from each grid and their posterior distribution, we performed the spectral analysis within a Bayesian framework.
The spectral characterization of the planet was carried out using the Python package \texttt{species}\footnote{\texttt{species} : \href{https://species.readthedocs.io/en/latest/}{species.readthedocs.io}} \citep{stolker_species_2023}, which was
developed for the atmospheric characterization of directly imaged exoplanets.
Additionally, we introduced the parameter A$_v$ to account for atmospheric extinction.
A$_v$ was parameterized by the $V$ -band extinction, considering the extinction law from \cite{cardelli_relationship_1989}.
We conducted the analysis both with and without the mid-infrared data points to assess the importance of the MIRI contribution. 
The first case is referred to as the free scenario, without any priors on the atmospheric parameters.
Second, we ran the analysis with priors on the surface gravity (prior g) with the goal to restrain the parameter space.
We chose a prior of $\mathrm{log} (g)$ = 3.9 $\pm$ 0.1 (according to the radius estimated from evolutionary models R = 1.35 $\pm$ 0.05 R$_{Jup}$ and the mass M = 4 $\pm$ 1 M$_{Jup}$ from previous studies; see Sect. \ref{sec:prop_planet}). This was also done recently by \cite{palma-bifani_atmospheric_2024} to obtain more accurate atmospheric fits with evolutionary models.
We finally ran an analysis with a prior for the radius of 1.35 $\pm$ 0.05 R$_{Jup}$ (prior R).
These three analyses are presented in Tab. \ref{tab:models_best_fits}, which summarizes the inferred parameters.
The fit bounds align with those of the atmospheric grids, spanning a radius ranging from 0.5 to 2.5 $R_{jup}$ and A$_v$ from 0 to 20.
The Bayesian analysis was run with 3000 live points with \texttt{MultiNest}.
We applied weighting factors based on the full width at half maximum of the filter profiles or the wavelength spacing calculated from the spectral resolution.
This prevented the fit from giving excessive importance to IFS with higher density measurements at the cost of the photometric points that cover a broader spectral range.

As we aim to explore whether a circumplanetry disk (CPD) surrounds the planet, a blackbody component was also included together with the atmospheric spectrum to account for the thermal emission that would come from an accretion disk. 
This emission is characterized by two free parameters: the effective temperature bounded between 100\,K and 2000\,K, and the
disk radius bounded between 1 and 1000\,R$_{Jup}$ \citep[similarly to the approach employed in][]{stolker_characterizing_2021}.
%

\begin{figure*}[h!]
    \centering
    \includegraphics[width=18cm]{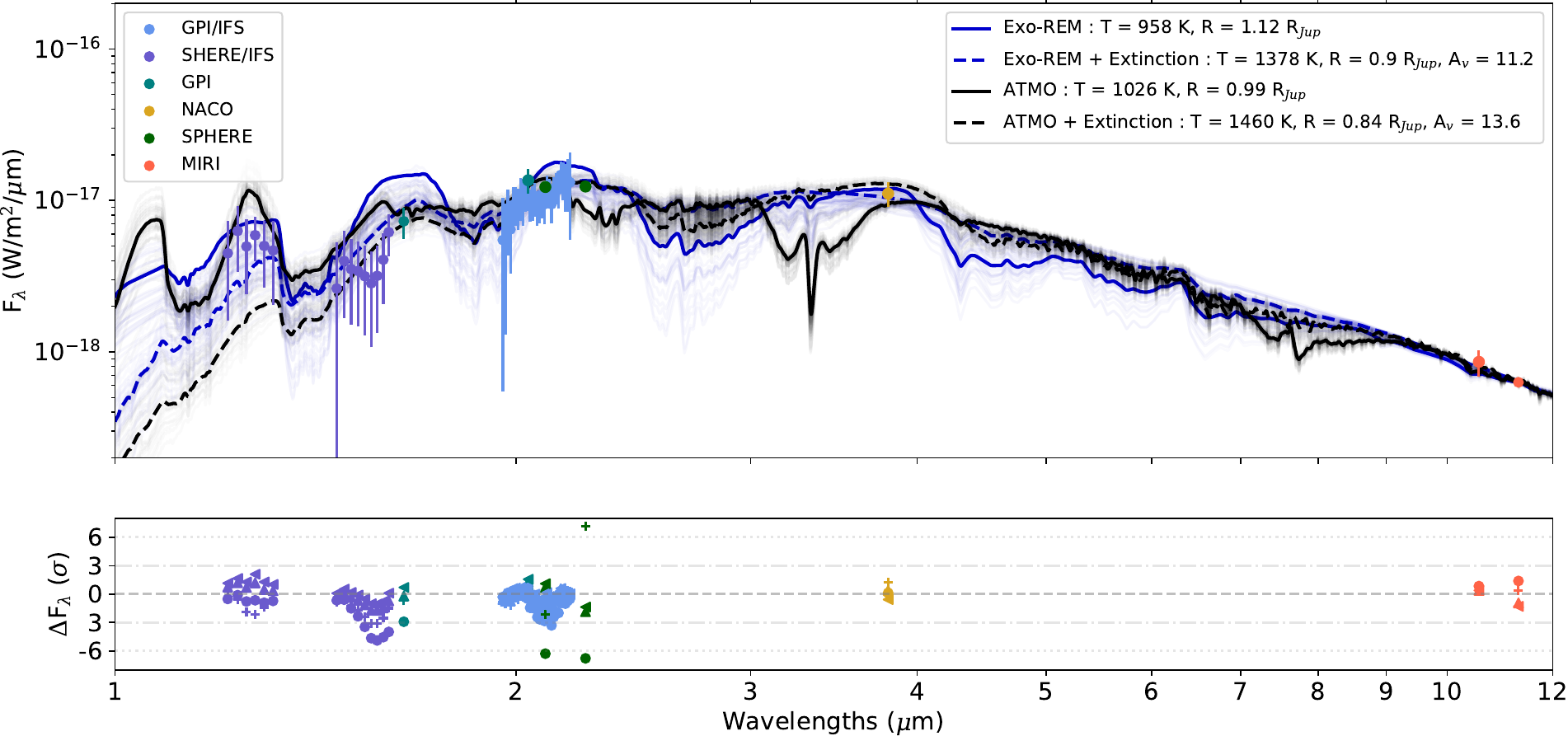}
    \caption{\texttt{Exo-REM} and \texttt{ATMO} best fits with all free parameters.
    The models with extinction are represented with dashed lines and those without extinction with plain lines. 
    The faint lines correspond to models from the posteriors distribution at $\pm$ 1 $\sigma$.
    The residuals between data and \texttt{Exo-REM} models without extinction are indicated with filled circles, and crosses show \texttt{ATMO} models. 
    The cases with extinction are indicated with upward triangles for \texttt{Exo-REM} and left triangles for \texttt{ATMO}. The same figure, adding a CPD component in the fitting process, is presented in Fig \ref{fig:Spectra_FM_CPD_HD95086}}.
    \label{fig:Spectra_FM_HD95086}
\end{figure*}

The best fits obtained with \texttt{Exo-REM} and \texttt{ATMO} are displayed in Fig. \ref{fig:Spectra_FM_HD95086} (free scenario).
%
The atmospheric parameters measured with both near- and mid-IR data are consistent with the values obtained when using only the near-IR data.
Tables \ref{tab:models_best_fits_nomiri} and Fig. \ref{fig:Spectra_FM_noMIRI_HD95086}
summarize the values measured when we fit only the near-IR data points.
The radius and temperature estimates (and consequently, the luminosity estimates) come with improved uncertainties when they are combined with mid-IR photometry.
The parameters derived from the two atmospheric models \texttt{Exo-REM} and \texttt{ATMO} agree within the error bars.
%
The C/O ratio is closer to solar values (C/O$_{solar}$ = 0.55) when atmospheric extinction is taken into account in the fit, otherwise, a subsolar C/O is favored (but with large uncertainties).
We measured a supersolar metallicity ([M/H] > 0) with \texttt{Exo-REM} and a subsolar metallicity with \texttt{ATMO} ([M/H] < 0) in most cases, again with large uncertainties. 
It is important to note that C/O and metallicity both affect the spectral shapes of the lines, and thus, we do not anticipate a precise estimation of these parameters using photometric data alone.
We note that the uncertainties become even broader when atmospheric extinction is added.
We measured two possible temperature ranges, depending on whether we took atmospheric extinction into account, but they almost overlap.
The radii measured are too small with respect to evolutionary model predictions for the system age; similarly, the masses measured are low due to unrealistically small surface gravity parameters (in the case of no atmospheric extinction).
When atmospheric extinction is added, $\mathrm{log} (g)$ and hence the masses are higher and more realistic, but the radius is lower.
We therefore added more priors before we ran the fitting process.

An additional prior to the surface gravity (prior g in Tab. \ref{tab:models_best_fits}) facilitates obtaining realistic mass measurements (between 2.3 and 4.2 M$_{Jup}$, depending on the models).
The effect on the radius (and all other parameters) is negligible within the uncertainties, however.

To continue, we ran a fit with a restrictive Gaussian prior on the radius value of R = 1.35 $\pm$ 0.05\,R$_{Jup}$.
This led to consistent radius and mass measurements with evolutionary models. 
In this case, \texttt{Exo-REM} resulted in lower $\mathrm{log} (g)$ and \texttt{ATMO} in higher and unrealistic $\mathrm{log} (g)$ compared to the case without prior information. 
With \texttt{Exo-REM}, an additional prior led to a lower temperature, even when we took an extinction parameter A$_v$ into account.
With  \texttt{ATMO}, the parameters were similar, but we note higher uncertainties on the surface gravity.\\
\begin{table*}[h!]
    \caption{Summary of the best-fit parameters in each case.}
    \centering
    \begin{tabular}{c|ccccccccccc}
        \hline
        \hline
         Model & T$_{eff}$ & log(g) & [Fe/H] & C/O & $\gamma$ & R (R$_{jup}$)& A$_v$ & log(L) & M (M$_{Jup}$)\\
         \hline
         \hline
         \texttt{Exo-REM} & \\
        free & 
        958+$^{+54}_{-56}$ & 3.33$^{+0.54}_{-0.21}$ & 0.53$^{+0.26}_{-0.29}$ & 0.34$^{+0.30}_{-0.17}$ & -- & 1.12$^{+0.07}_{-0.07}$ & -- & -5.00$^{+0.03}_{-0.05}$& 1.1$^{+2.8}_{-0.4}$\\

        prior g & 936+$^{+66}_{-63}$ & 3.90$^{+0.1}_{-0.1}$ & 0.55$^{+0.22}_{-0.19}$ & 0.34$^{+0.30}_{-0.12}$ & -- & 1.14$^{+0.08}_{-0.08}$ & -- & -5.02$^{+0.04}_{-0.06}$& 4.1$^{+1.2}_{-1.0}$\\

        prior R & 
        834+$^{+26}_{-29}$ & 3.42$^{+0.57}_{-0.29}$ & 0.85$^{+0.10}_{-0.19}$ & 0.23$^{+0.16}_{-0.08}$ & -- & 1.30$^{+0.04}_{-0.04}$ & -- & -5.11$^{+0.03}_{-0.04}$& 1.8$^{+4.7}_{-0.9}$\\

        \hline 
        free & 
        1378$^{+243}_{-183}$ & 4.11$^{+0.60}_{-0.70}$ & 0.15$^{+0.53}_{-0.42}$ & 0.48$^{+0.21}_{-0.23}$ & -- & 0.90$^{+0.05}_{-0.04}$ & 11.2$^{+4.7}_{-4.0}$ & -4.56$^{+0.25}_{-0.21}$ & 4.2$^{+11.9}_{-3.1}$\\

        prior g & 1341$^{+236}_{-174}$ & 3.90$^{+0.1}_{-0.1}$ & 0.19$^{+0.54}_{-0.47}$ & 0.48$^{+0.21}_{-0.25}$ & -- & 0.90$^{+0.06}_{-0.03}$ & 10.4$^{+4.9}_{-3.8}$ & -4.6$^{+0.26}_{-0.2}$ & 2.7$^{+0.7}_{-0.5}$\\

        prior R & 
        857$^{+37}_{-40}$ & 3.74$^{+0.56}_{-0.51}$ & 0.87$^{+0.09}_{-0.18}$ & 0.22$^{+0.20}_{-0.08}$ & -- & 1.28$^{+0.05}_{-0.04}$ & 1.41$^{+2.1}_{-1}$ & -5.08$^{+0.05}_{-0.05}$ & 3.7$^{+9.3}_{-2.5}$\\
        
         \hline 
         \hline 
         \texttt{ATMO} & \\
               
        free & 1026+$^{+21}_{-22}$ & 3.06$^{+0.09}_{-0.04}$ & -0.27$^{+0.46}_{-0.16}$ & 0.34$^{+0.06}_{-0.03}$ & 1.02$^{+0.02}_{-0.01}$ & 0.99$^{+0.02}_{-0.02}$ & -- & -5.0$^{+0.02}_{-0.03}$& 0.5$^{+0.1}_{-0.1}$\\

        prior g & 987+$^{+22}_{-24}$ & 3.87$^{+0.1}_{-0.1}$ & -0.46$^{+0.06}_{-0.03}$ & 0.33$^{+0.05}_{-0.02}$ & 1.02$^{+0.01}_{-0.0}$ & 1.05$^{+0.03}_{-0.03}$ & -- & -5.0$^{+0.02}_{-0.02}$& 3.3$^{+0.9}_{-0.7}$\\

        prior R & 884+$^{+23}_{-20}$ & 4.48$^{+0.21}_{-0.23}$ & -0.45$^{+0.08}_{-0.04}$ & 0.34$^{+0.06}_{-0.03}$ & 1.02$^{+0.01}_{-0.01}$ & 1.30$^{+0.04}_{-0.05}$ & -- & -5.0$^{+0.02}_{-0.02}$& 46.0$^{+33.7}_{-20.6}$\\

        \hline
        free & 1460$^{+219}_{-172}$ & 4.07$^{+0.78}_{-0.70}$ & -0.12$^{+0.35}_{-0.25}$ & 0.49$^{+0.13}_{-0.13}$ & 1.03$^{+0.01}_{-0.01}$ & 0.84$^{+0.06}_{-0.06}$ & 13.6$^{+3.8}_{-3.7}$ & -4.52$^{+0.18}_{-0.16}$ & 3.3$^{+17.4}_{-2.7}$\\

        prior g & 1438$^{+249}_{-164}$ & 3.90$^{+0.1}_{-0.1}$ & -0.12$^{+0.34}_{-0.26}$ & 0.50$^{+0.13}_{-0.13}$ & 1.03$^{+0.01}_{-0.01}$ & 0.83$^{+0.07}_{-0.07}$ & 12.9$^{+4.3}_{-3.5}$ & -4.55$^{+0.2}_{-0.15}$ & 2.2$^{+0.7}_{-0.6}$\\

        prior R & 1349$^{+185}_{-111}$ & 4.38$^{+0.60}_{-0.63}$ & 0.49$^{+0.01}_{-0.01}$ & 0.69$^{+0.01}_{-0.01}$ & 1.03$^{+0.01}_{-0.01}$ & 1.31$^{+0.04}_{-0.04}$ & 12.1$^{+4.0}_{-3.0}$ & -4.27$^{+0.22}_{-0.15}$ & 16.47$^{+49.1}_{-12.9}$\\
        
        \hline
         
    \end{tabular}
    \tablefoot{
    The free case refers to no priors and corresponds to Fig. \ref{fig:Spectra_FM_HD95086}, prior g indicates the prior on the surface gravity, and prior R indicates the prior on the radius from the Bayesian analysis. 
    The two sets of solutions correspond to the parameters measured with and without the extinction parameter (A$_v$) in models \texttt{ATMO} and \texttt{Exo-REM}.}
    \label{tab:models_best_fits}
\end{table*}

The results with the CPD contribution are summarized in the appendix in Tab. \ref{tab:models_best_fits_CPD}, and Fig. \ref{fig:Spectra_FM_CPD_HD95086} presents the best fits
(see Tab. \ref{tab:models_best_fits_CPD_noMIRI} and Fig. \ref{fig:Spectra_FM_noMIRI_CPD_HD95086} for the same analysis with the near-IR data alone).
Additional near-IR extinction (A$_v$) can also be added with the CPD hypothesis.
The parameters measured for the CPD are T$_{bb}$ $\sim$ 130--140\,K and a radius of $\sim$ 15--20\,R$_{Jup}$, depending on the atmospheric models. These values are consistent with \cite{desgrange_-depth_2022} for the temperature, but not for the radius, which is higher in our analysis.
The atmospheric parameters measured under the assumption of a CPD component have more significant uncertainties.
None of the values we measured has any significance for the atmospheric fit.
Adding this contribution results in an even lower radius for the planets. It is therefore not consistent with evolutionary models and is a poorer fit.
In conclusion, the mid-IR data allow us to rule out the hypothesis of a warm CPD, but they are not sufficient to definitely exclude a cold debris disk around the planet that would emit at even longer wavelengths.

\section{Debris disks}
\label{sec:debris_disks}
\subsection{Inner disk}
\label{sec:inner_disk}
\begin{figure*}[h!]
    \centering
    \includegraphics[width=18cm]{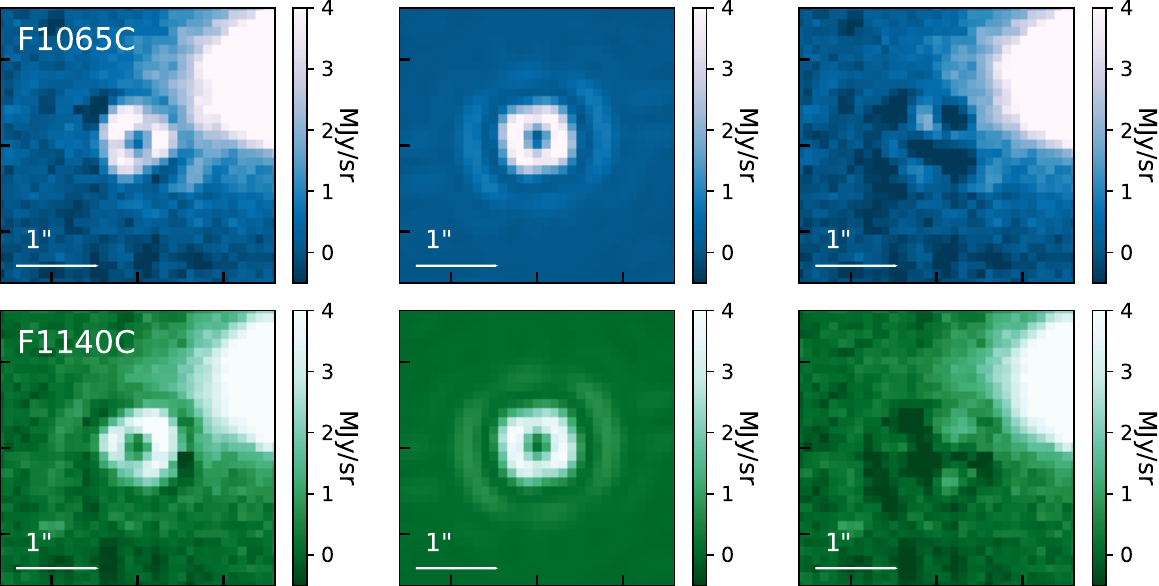}
    \caption{Inner disk modeling. 
    Left: Data after subtraction of the best-fit planet model PSF, corresponding to the right panel from \ref{fig:image_sub}.
    Middle: Best-fit model of the inner disk. 
    Right: Residual data subtracted by the best-fit disk model.
    The top panel corresponds to the F1065C, and the bottom panel corresponds to the F1140C.}
    \label{fig:model_disk}
\end{figure*}

The central region of each image is dominated by the emission of the inner disk component of the system. 
The inner part of the system was studied using photometry by \cite{su_debris_2015} and \cite{su_alma_2017}, who inferred a warm belt located around 7 to 10\,au with a temperature of 187 $\pm$ 26\,K.
An even warmer dust belt (300\,K) was suggested at a closer distances (2\,au).
We modeled the warm component of the disk by assuming a uniform face-on disk model, involving a single parameter, its radius, which ranged from 2 to 50\,au.
We used the diffraction model from \cite{boccaletti_fast-moving_2015}, calibrated with commissioning data.
We observe similar results to the inner disk of HR\,8799 as observed at F1550C \citep[see Fig. 5 from][]{boccaletti_imaging_2024}.
At F1065C and F1140C, the disk is unresolved for a radius smaller than 5\,au, and the image resembles that of a coronagraphic stellar image of a typical point source.
Between 10 and 30\,au, we obtain an image similar to the one observed in the MIRI dataset.
Finally, beyond 30\,au, we recover the coronographic image of an extended source (in which the transitions of the 4QPM are visible as shadows).
To determine the disk radius from the data, we minimized the residuals between the data and the disk models (spanning a range of disk sizes) within a region with a radius of 2$''$, centered on the coronagraphic mask.
Additionally, a circular region with a radius of 1 $\lambda/D$ was used to hide the planet to ensure that it does not impact the minimization process.
We included two free parameters to adjust the position of the model (in x and y), which may not be perfectly centered with respect to the data.
We find that the optimal disk model has a radius of 25\,au in both filters F1065C and F1140C.
However, all models spanning from 10 to 30\,au yield comparable results 
(residuals not exceeding 2.1\% of the optimal case).
Figure \ref{fig:model_disk} displays the best-fit disk models for both filters.
The residuals obtained by subtracting the model from the data clearly show the asymmetries observed in the original data.
These asymmetries cannot be accounted for by an offset on the coronagraph \citep{boccaletti_imaging_2024}.
Either they are real density variations of the inner disk, or they are artifacts generated by imperfections at the $\muup$m scale on the 4QPM surface, which induce local phase variations.
%
We note that a joint modeling of the planet and disk would require a more accurate disk model that takes the asymmetries into account.
A deeper analysis like this is beyond the scope of this paper, but would be necessary to better understand how the MIRI coronagraphs impact the structure of inner disk emissions.

The flux density of the inner disk was estimated based on a disk model and the corresponding coronagraph attenuation.
We measure 2.8\,mJy, 1.67\,mJy and 1.25\,mJy at F1056C for disk radii of 10, 20 and 30\,au, respectively.
In filter F1140C, the fluxes amount to 2.6\,mJy, 1.85\,mJy, and 1.11\,mJy.
Based on the SED, \cite{su_debris_2015} inferred a flux of $\sim$5.2\,mJy at 10.5\,$\mu$m and 6.3\,mJy at 11.3\,$\mu$m. 
They argued that the warm component of the SED is produced by an independent inner belt between 7 and 10\,au.
The SED modeling based on Spitzer/IRS data provides twice higher flux values for the inner disk component and is better aligned with the values derived from MIRI data when we assume a disk size of  10\,au.
However, based on the MIRI images, we cannot exclude that the disk lies at a more distant location from the star, and therefore, that its flux is lower.
Furthermore, we note that based on the spatial resolution from Spitzer/IRS ($\sim2''$, in comparison with $\sim0.3''$ for JWST), it is likely that the SED measured previously
also integrates the signal from the background objects.
Therefore, the previous flux values might be overestimated.
Along the same line, the SED modelling also identifies an IR excess shortward of 10 $\mu$m. 
\cite{su_debris_2015} suggested that either there is an innermost hotter component emission, which would correspond to a 300\,K blackbody, or the emission from the warm component is a weak silicate feature, which remains undetected so far.
The MIRI observations, which provide a higher angular resolution than Spitzer, would argue that this discrepancy is caused by background objects.

\subsection{Outer disk}
\label{sec:outer_disk}
\begin{figure}[h]
    \centering
    \includegraphics[width=8.8cm]{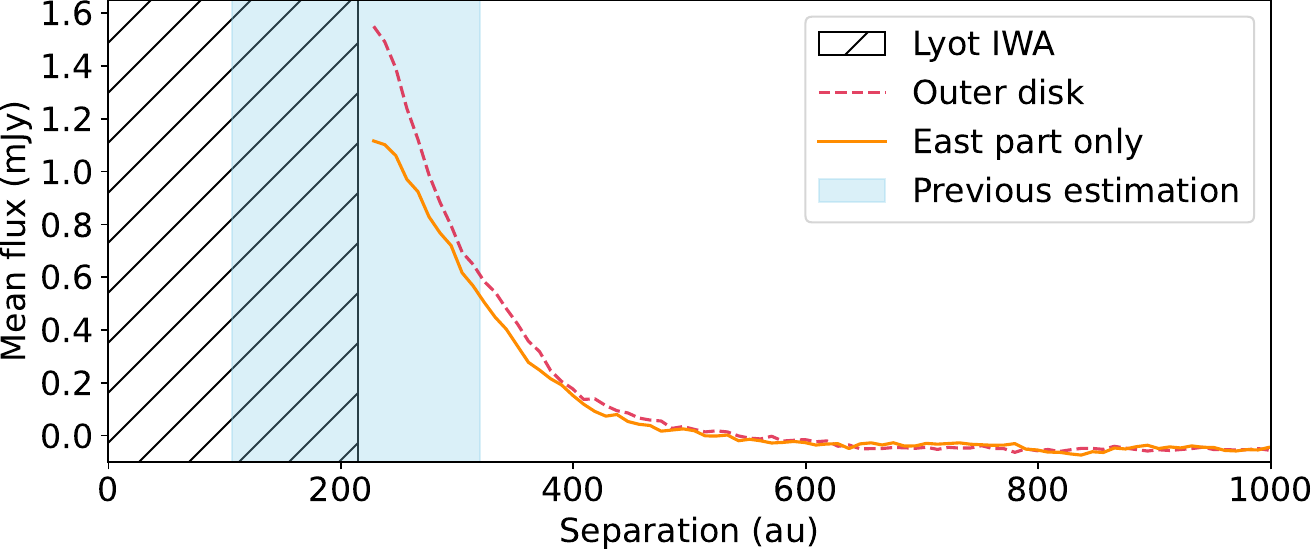}
    \caption{Average flux in F2300C as a function of the separation (indicated in astronomical units).
    The shaded region in blue comes from ALMA data \citep{su_alma_2017}.}
    \label{fig:outer_disk}
\end{figure}

The outer disk is clearly detected at 23 $\mu$m. The extent of its emission is consistent, if slightly more extended, with the ALMA data \citep[measured from 106 $\pm$ 6 au and 320 $\pm$ 20 au from the star in][]{su_alma_2017}. 
Figure \ref{fig:outer_disk} presents the average emission of the outer disk as a function of its separation from the host star, and  it extends up to 400\,au in the MIRI image.
We distinguish the average flux over the whole image (dashed red) and in the eastern part (orange), as the flux from the outer disk in the western part is heavily contaminated by background galaxies.
The inner working angle (IWA) of MIRI coronagraphs is indicated in black. It corresponds to the smallest observable separation ($\pm$ 2.5$''$ for the F2300C, i.e. 216.2 au). Therefore, we only measured the emission of the disk above this value.

\section{Discussion}
\label{sec:discusion}
\subsection{Properties of the planet}
\label{sec:prop_planet}
The atmospheric analysis of the planet we presented in Sect. \ref{sec:forward_modeling} without any prior (free) or extinction resulted in a luminosity of log\,L = -5.0 $\pm$ 0.05 (Tab. \ref{tab:models_best_fits})
 for both \texttt{Exo-REM} and \texttt{ATMO}. 
To obtain an independent measurement of the radius, we also used the \texttt{ATMO} evolutionary models \citep{phillips_new_2020}.
When the system is considered to be part of the Carina association with the younger age assumption of 13.3\,Myr \citep{booth_age_2021}, the luminosity measurement from the atmospheric modeling corresponds to a mass of 3.63\,M$_{Jup}$ and a radius of 1.33\,R$_{Jup}$. 
For an older age of 41 Myr, we obtain a mass of 6.74\,M$_{Jup}$ and a radius of 1.24\,R$_{Jup}$.
Finally, if HD\,95086\,b belonged to the MELANGE-4 associations at 27\, Myr \citep{wood_tess_2023}, we would infer a mass of 5.37\,M$_{Jup}$ and radius of 1.28\,R$_{Jup}$.
When atmospheric extinction is taken into account in the atmospheric fit, the luminosity increases to log\,L = -4.5 $\pm$ 0.2,
which translates into a mass of 8.75 M$_{Jup}$ and a radius of 1.3 R$_{Jup}$ (assuming the system age to be 27 Myr).
These predicted radius values with evolutionary models remain higher than the values obtained through atmospheric fitting.
Even with constraints on the surface gravity using priors, the fitting process still converges toward unusually small radii.
This discrepancy is likely attributable to dusty and cloudy atmospheres.
The radius values, which are more closely aligned with the evolutionary models, are observed when no additional atmospheric extinction or circumplanetary dust is taken into account in the fitting process.

Compared to the findings in \cite{desgrange_-depth_2022}, MIRI photometry allows us to rule out the model from \cite{madhusudhan_model_2011}, who assumed forsterite clouds because the model predicts fluxes that are too high at mid-infrared wavelengths.
MIRI photometry provides results that agree with one of their hypotheses: lower temperatures (800 -- 1300 K), a small to medium amount of extinction (A$_v$ $\leq$  10 mag), and an atmosphere with a supersolar metallicity.
This last is more consistent with the \texttt{Exo-REM} models.

Overall, mid-infrared photometry leads to improved measurements of the luminosity and temperature. 
We achieved an increased precision in the atmospheric parameters of the planet, which reduced the uncertainties for several parameters significantly, for instance, for the luminosity, for which the uncertainties are reduced by a factor of $\sim$ 3. 
The uncertainties on the radius are decreased by a factor 2 to 6 (depending on the atmospheric model used).
As presented in Tab.\ref{tab:models_best_fits} and Tab. \ref{tab:models_best_fits_nomiri}, the uncertainties on the luminosity are $\sim \pm$ 0.1 with near-IR values alone and decrease to $\sim \pm$ 0.04 (e.g., with \texttt{Exo-REM} models and the free scenario).

For young giant exoplanets with a dusty and/or cloudy atmosphere, it remains challenging to obtain atmospheric fits that align with evolutionary model predictions.
We can hypothesize that some ingredients might be missing in the models: clouds, dust, or others.
Concerning the metallicity and the C/O ratio, we note that low-resolution spectra and photometry are insufficient to constrain either of these parameters, which primarily affect the strength of the lines. 
As a consequence, we obtained discrepancies depending on the atmospheric model.
The measured solar C/O ratio would be consistent with the formation through gravitational instability \citep{boss_giant_1997} ; but according to the core-accretion models, the solar to subsolar C/O ratios can indicate that its atmosphere has been contaminated by evaporating planetesimals \citep{oberg_effects_2011}.
However, due to the substantial uncertainties on the C/O ratio, we refrain from using this parameter for drawing conclusions about the planet formation.
Likewise, the measured metallicity is contingent on the models that are used and remains unconstrained.
The molecule NH$_3$ could have been detected by comparing the flux measured at F1065C and F1140C \citep{danielski_atmospheric_2018}. 
We do not detect this molecule in the atmosphere of HD\,9086\,b. According to the \texttt{Exo-REM} models, if NH$_3$ is indeed present, its abundance at these temperatures would be very low. 
As a conclusion, the nondetection of NH$_3$ is consistent with atmosphere model predictions.

A higher resolution spectrum is needed to overcome the degeneracies that remain in the atmospheric parameters and to measure molecular composition and the impact of clouds.
This could be provided, for example, with NIRSpec or MIRI/MRS at mid-infrared wavelengths.
Although these instrument modes are not equipped with coronagraphs, the use of stellar PSF subtraction methods has proved difficult but possible,
at least in the case of the bright planet $\beta$\,Pic\,b \citep{worthen_miri_2024}.
The application of PSF subtraction methods along with molecular mapping techniques could unveil the presence of molecules in this atmosphere \citep{malin_simulated_2023}.

\subsection{Circumplanetary disk}
Based on mid-IR photometry, we also indicate that the CPD hypothesis is unlikely. Only an extended cold ($\sim$ 100\,K) CPD would be consistent with the mid-IR data, but longer-wavelength data are required to confirm this hypothesis \citep{benisty_circumplanetary_2021}.
An alternative hypothesis presented by \cite{desgrange_-depth_2022} suggested that a fraction of the atmospheric flux is obscured by circumplanetary dust.
This could account for the lower radius determined through atmospheric modeling when compared to predictions from evolutionary models.
However, the authors noted that the near-infrared spectrum alone makes it challenging to determine the specific location of the dust, whether it is situated in the upper atmosphere or around the planet.
To evaluate the impact of a hypothetical CPD, \cite{desgrange_-depth_2022} used two methods for modeling the additional CPD component.
The first was a circumplanetary primary viscous disk in which the planet still accreted material \citep{zhu_accreting_2015} . They concluded, however, that this model leads to a poorer fit to the near-IR data.
Viscous disk with accreting planets have been identified in younger system than HD\,95086, such as PDS\,70\,b at an age of $\sim$ 5 Myr \citep{christiaens_evidence_2019}.
This scenario is not favored for HD\,95086 because it is an older system 
(even with the youngest hypothesis of $\sim$ 13.3 Myr for the system)
with a lower amount of CO \citep{booth_deep_2019}.
Therefore, we only tested the alternative model of the CPD here, which is the following: The dust is located close enough
to the planet to be heated to high enough temperatures so that an IR excess is produced. 
This was modeled with a blackbody component added to the atmospheric emission of the photosphere.
We find that adding this component in the fitting process leads to a poorer fit and provides a negligible contribution to the model.

Finally, \cite{chen_observability_2022} showed that mid-IR wavelengths (and longward) are the best range to search for CPDs:  if there were a warm CPD around this planet, it would have been observable with JWST/MIRI.
A broader wavelength coverage would allow us to draw a more comprehensive conclusion about the potential presence of a cold CPD around the planet.

\subsection{Challenges inherent to the inner disk detection}

The inner disk separation was estimated to be  7 -- 10 au from the star \citep{su_debris_2015}. 
It should therefore not be resolved with the angular resolution capability of MIRI ($\lambda/D$ corresponds to $\sim 30$\,au). 
The detection of the inner disk was therefore not expected. 
Similarly to HR\,8799 \citep[detected at F1550C,][]{boccaletti_imaging_2024}, however, the detection of the inner disk is the result of a diffraction leakage due to the extreme sensitivity of the MIRI coronagraphs at small angles.

Even though HD\,95086 is more distant, the inner disks from both systems have similar flux levels, and the inner disk of HD\,95086 is slightly warmer than that of HR\,8799 \citep[175\,K as opposed to 150\,K, ][]{su_debris_2009, su_debris_2015}.
Furthermore, the inner disk of HR\,8799 is fainter than the four giant planets, in contrast to the HD\,95086 system, whose planet is fainter than the inner disk in the F1065C and F1140C filters.

One issue is related to the fact that the techniques for stellar subtraction, such as PCA and optimized linear subtraction, are not as effective as they were in the context of ground-based near-infrared direct imaging of exoplanets.
We used a simpler method for subtracting the stellar contribution here. However, accounting for the disk and planet contributions jointly in the stellar subtraction process could lead to improvements.
Future developments of reference libraries and advancements in algorithms could provide better stellar subtraction.
Moreover, the asymmetries observed in the inner disk (Fig. \ref{fig:model_disk}) might also be attributed to an imperfect subtraction of the stellar diffraction, as illustrated in the appendix \ref{app:other_red}.

Due to these asymmetries, we observe that the planet PSF overlaps even more with the disk in the F1065C than in the F1140C.
This poses a greater challenge to accurately distinguish the contribution of the planet from that of the inner disk.
Even if the disk is masked to fit the planet PSF in order to extract the photometry, and the planet seems well subtracted from the data (Fig. \ref{fig:planet_model}), it is still possible that the planet photometry is overestimated because of the disk. 
Most atmospheric models result in a higher residual in F1065C than in the F1140C filters. 
This could result from atmospheric features that are not considered in the model, but also from a slight overestimation of the planet flux.
An improved disk modeling would allow us to jointly capture the planet PSF and the disk, thus confirming the planet photometry.
Finally, modeling the inner disk represents a challenge inherent to the 4QPM, the shape of the disk will strongly depend on its position with respect to the center of the coronagraph.
The model has to take into account 4QPM transmission that is strongly nonlinear at these separations. 

\subsection{Additional planets}
\label{sec:add_planets}
\begin{figure}
    \centering
    \includegraphics[width=9cm]{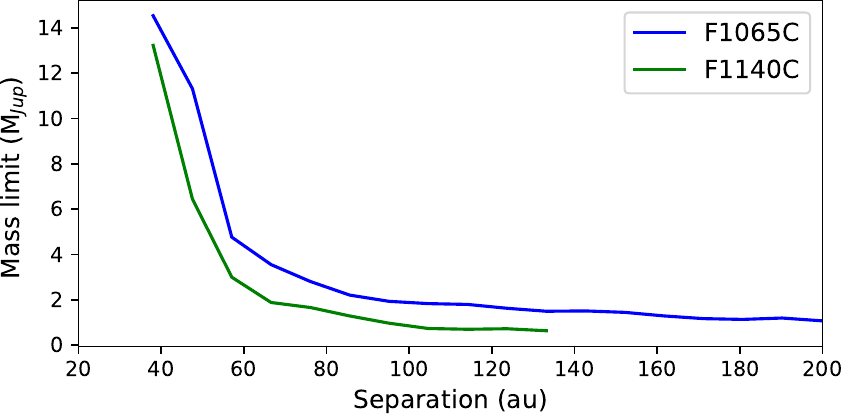}
    \caption{Sensitivity in mass at each separation for the HD\,95086 system observed with JWST/MIRI.}
    \label{fig:mass_limit}
\end{figure}

There are multiple ways to explore additional planets.
We can directly evaluate the detection performance from MIRI data or consider dynamical arguments such as the locations of the belts.
The multiple-belt architecture of the system provides compelling evidence for the existence of additional planets.
The clearing of a disk is usually thought to be driven by the gravitational effect of giant planets, which influence the system architecture, as shown for PDS\,70 and HR\,8799, where planets are located in a gas-depleted cavity or in between two debris belts.
Similarly, for HD\,95086, the gap between the inner warm and the cold outer belt is too wide to be accounted for by a single planet.
Many scenarios have been explored to infer the positions and masses of additional undetected planets \citep{su_debris_2015}.
The single-planet scenario would imply a high eccentricity for planet b, but this was refuted by \cite{rameau_constraints_2016}. This confirms that a multiple-planet scenario is required to explain the dynamical stability of the system.
\cite{su_debris_2015} showed that an equal-mass ($\sim$ 5 M$_{Jup}$) four-planet configuration of geometrically spaced orbits could maintain the gap between the warm and cold debris disks. 
However, no additional planets have been discovered with the most efficient high-contrast imaging instruments such as VLT/SPHERE and Gemini/GPI.
Based on dynamical arguments for the clearing zone, \cite{chauvin_investigating_2018} derived a minimum mass in the cavity of 0.35 M$_{Jup}$ when only two or three giants planets are considered, and depending on their respective separations.
Comparing these theoretical results to HARPS and SPHERE detection limits, \cite{chauvin_investigating_2018}
noted that the system is more likely to have two additional planets in the cavity, with typical masses of 0.35 to 6 M$_{Jup}$ for a semimajor axis of 10 -- 30\,au ; or 5 M$_{Jup}$ beyond 30\,au.
Using the K-Stacker algorithm \citep{le_coroller_k-stacker_2020} and using many post-processing algorithms, \cite{desgrange_-depth_2022} pushed the detection performances even further. 
Although no additional planets were detected, they ruled out the existence of any planets above 5 M$_{Jup}$ in the system located at a distance greater than 17\,au (at the 50\% confidence level).
At the distance of HD\,95086 (86.46\,pc), the MIRI  spatial resolution is 28.5\,au.
Considering the inner disk contribution at MIRI wavelengths and the spatial resolution of MIRI, new detections below 50\,au are unlikely, with planet HD\,95086\,b already at the shortest detectable distance.
In conclusion, we cannot constrain any innermost planets with these observations.
In the outermost regions of the system, MIRI has the capability to detect lower-mass and colder planets.  \cite{carter_direct_2021} measured the  detection limits for HIP\,65426\,b (at 108\,pc) to derive a minimum mass of 1.5 M$_{Jup}$ ($\sim$ 550\,K) from $\sim$ 150 to 2000 au.
We measured the contrast curves for the two MIRI images at F1065C and F1140C. We took the coronagraph transmission into account and measured the attenuation at several separations using simulated PSF with \texttt{WebbPSF}. These PSFs were simulated with a position angle of 45$^\circ$, which provides the most optimistic coronagraphic throughput. The detection limits are poorest close to the edge of the quadrants.
Assuming the synthetic flux of the star in MIRI filters, we converted the contrast into masses based on \texttt{ATMO} evolutionary model predictions.
This provided the mass limits that are accessible for the HD\,95086 system with JWST/MIRI, as displayed Fig. \ref{fig:mass_limit}. We can exclude at 5 $\sigma$ any planets more massive than 2.6 M$_{Jup}$ at a separation larger than 60\,au and 0.8 M$_{Jup}$ at a separation larger than 100\,au. 
However, planets at larger separations are not particularly expected, as the outer disk starts at $\sim$ 100 au.

While we analyzed the processed image at F1140C, we identified a suspicious feature that is located farther away from planet b (1.53$''$ equivalent to $\sim$ 132 au in projected physical separation, i.e., within the outer disk) with a similar position angle. The feature is faint, but appears to be point-like. 
We conducted tests to verify whether it was an actual source or an artifact.
First, considering the photometry of this feature, it would align with an atmospheric model at 500\,K and 1.2\,R$_{Jup}$ (or a slightly higher temperature and accordingly lower radius).
If this is the case, we could expect a nondetection in the F1065C filter due to NH$_3$ absorption. However, the brightness measured at F1140C suggests a detectable planet at F1065C, unless it exhibits more significant NH$_3$ absorption than predicted by the \texttt{Exo-REM} model (see Fig. \ref{fig:photom_new_source} in the appendix).
%
Second, the modeling of the PSF for this point source is unsuccessful 
(see Fig. \ref{fig:model_new_source} in the appendix) because its shape does not resemble that of a typical PSF.
Finally, in the raw data, this point source is located at the position of a bad pixel.
It can therefore likely be attributed to ineffective correction of bad pixels.
The point source also appears differently in the \texttt{spaceklip} data reduction, which handles bad pixels differently.
We conclude that a planetary nature for this possible point source is unlikely.

\subsection{A young analog of the well-studied system HR\,8799}

The system HD\,95086 exhibits notable similarities with other systems that have imaged planets.
It has been referred to as a younger analog to the well-studied HR\,8799 \citep{su_debris_2015}, which was also observed with MIRI coronagraphs \citep{boccaletti_imaging_2024}.
Both systems have a warm excess near the water-ice line and a cold excess surrounded by an extended outer belt.
The inner disk contribution has been observed for the first time in both systems owing to the capabilities of MIRI coronagraphs. 
For the HD\,95086 system, it was observed in the F1065C and F1140C, and it is brighter than the planet, but for HR\,8799, the planets are brighter at these wavelengths and the inner disk is only dominant in the F1550C observation. 
This can be consistent with HD\,95086 being a younger system, but we cannot rule out that the inner disk of HR\,8799 is more difficult to detect because the four planets are too bright and overwhelm the image, even when the planets are located farther out.
The extended outer disk halos of both systems have a similar size, as confirmed with MIRI observations in the F2300C for HD\,95098, which is at least five times brighter. 
Both systems fall within the age range of 10 to 100 Myr. During this period, processes such as dynamical settling, the formation of rocky planets, and possibly still the formation of ice giants are expected to occur in planetary systems.
One of the main differences between the two systems arises from the number of discovered planets: HD\,95086 is twice more distant, which poses greater observational challenges, particularly if these planets are less massive, as suggested by theoretical expectations.
%

\section{Conclusion and perspectives}
\label{sec:conclusion}
We presented the MIRI coronagraphic images of the system HD\,95086, which is one of the first planetary systems that was observed at mid-infrared wavelengths.
\begin{itemize}
    \item We presented the first direct detection of the inner disk of the HD\,95086 system and detected the planet at both observation wavelengths.
    \item 
    With these MIRI observations, we exemplified some inherent challenges to mid-infrared high-contrast imaging.
    The contribution from the inner warm disk is detected, and background galaxies can be dominant at these wavelengths. This prevented us from using traditional stellar subtraction methods.
    Therefore, we used an optimized method to remove stellar diffraction without being impacted by the inner disk or any other background object.
    \item We extracted the photometry of planet HD\,95086\,b at 10.5 and 11.3 $\mu$m using various methods of flux normalization. 
    We verified that the latest \texttt{JWST} flux calibrations agree with the method based on measuring the contrast relative to the host star. 
   We observed variations of 13\% at F1065C and 2\% at F1140C between the flux calibration methods, and the flux uncertainties due to the inner disk contamination are 14\% at F1065C and 12\% at F1140C.
    \item   The uncertainties on the atmospheric parameter measurements are improved by a combination of the near-IR data and mid-IR photometry.
    The uncertainties on the luminosity measurement are reduced by at least a factor 2, and the uncertainties on the radius are reduced by a factor 3 to 7 (depending on the models and fitting hypothesis).
    \item The measured temperature of the planet ranges from 800 -- 1050\,K depending on the atmospheric model. 
    When additional extinction is taken into account in the forward modeling, we evaluate A$_v$ $\sim$ 10 -- 14, and therefore, we measure a higher temperature of 1340 -- 1460\,K.
    \item 
    Adding the mid-infrared information, we measured radius values ranging from $\sim$ 1 -- 1.14 R$_{Jup}$, which is closer to those derived with evolutionary models (1.24 -- 1.28 R$_{Jup}$), but they still do not perfectly align.
    \item Data with a higher spectral resolution are needed to constrain the atmospheric composition better, such as measuring a precise metallicity or atomic ratio that can be linked to the planetary formation.
    \item 
    We discard the previous hypothesis of a warm circumplanetary disk as a way of accounting 
    for the red near-infrared colors of the planet.
    %
    \item The outer disk belt is detected at 23$\mu$m. Its flux and size measurements are consistent with the previous detection at longer wavelengths.
    \item Asymmetries are clearly visible in the inner disk component, but it is challenging to draw conclusions about their origins because the effect of stellar subtraction is strong and is particularly pronounced at separations close to the center of the coronagraph.
    \item However, the inner disk is compatible with a coronagraphic image model with a radius of $\sim$ 10 to 30 au, 
    which is reasonably consistent with previous values, which placed it between 7 and 10 au.
    \item We measure a lower flux value for the inner disk in comparison to Spitzer. This is likely due to background objects that were unresolved with the Spitzer angular resolution.
    \item The hypothesized presence of an innermost belt (at $\sim$ 2 au) in previous studies may be attributed to the contamination of the SED by background objects.
\end{itemize}

Similarly to the HR\,8799 system, the young system HD\,95086 is a benchmark object on which to study the formation and evolution of planetary systems.
Follow-up observations with MIRI/MRS would be highly beneficial for this system.
As presented in Sect. \ref{sec:prop_planet}, the extraction of a medium-resolution spectrum would be challenging but definitive regarding the atmospheric properties of the planet, including the cold CPD hypothesis. 
Even without a spectral extraction of the planet, molecular mapping can be used to detect molecular species in the atmosphere of HD\,95086.
Moreover, although it is more challenging to detect the planet at the longer wavelengths of MIRI/MRS, it should be possible to measure an IR excess corresponding to an eventual CPD as compared to the expectation from atmospheric models.
Drawing conclusions about the composition of the inner disks based on these coronagraphic observations is difficult, but MIRI/MRS also has the potential to unveil their composition \citep[such as for PDS\,70,][]{perotti_water_2023}. 

\begin{acknowledgement}
This work is based on observations made with the NASA/ESA/CSA James Webb Space Telescope. The data were obtained from the Mikulski Archive for Space Telescopes at the Space Telescope Science Institute, which is operated by the Association of Universities for Research in Astronomy, Inc., under NASA contract NAS 5-03127 for JWST. These observations are associated with program \#1277.
Part of this work was carried out at the Jet Propulsion Laboratory, California Institute of Technology, under contrast with NASA (80NM0018D0004).
This publication makes use of VOSA, developed under the Spanish Virtual Observatory (https://svo.cab.inta-csic.es) project funded by MCIN/AEI/10.13039/501100011033/ through grant PID2020-112949GB-I00. 
VOSA has been partially updated by using funding from the European Union's Horizon 2020 Research and Innovation Programme, under Grant Agreement nº 776403 (EXOPLANETS-A)
This work has made use of data from the European Space Agency (ESA) mission
{\it Gaia} (\url{https://www.cosmos.esa.int/gaia}), processed by the {\it Gaia}
Data Processing and Analysis Consortium (DPAC,
\url{https://www.cosmos.esa.int/web/gaia/dpac/consortium}). Funding for the DPAC
has been provided by national institutions, in particular the institutions
participating in the {\it Gaia} Multilateral Agreement.
This research has made use of the VizieR catalogue access tool, CDS, Strasbourg, France (DOI : 10.26093/cds/vizier). The original description of the VizieR service was published in 2000, A\&AS 143, 23
Software: python, astropy, numpy, scipy  matplotlib.
M.M., A.B., P.-O.L, C.C. acknowledges funding support by CNES.
B.V. thanks the European Space Agency (ESA) and the Belgian Federal Science Policy Office (BELSPO) for their support in the framework of the PRODEX Programme.
O.A. is a Senior Research Associate of the Fonds de la Recherche Scientifique – FNRS. OA thanks the European Space Agency (ESA) and the Belgian Federal Science Policy Office (BELSPO) for their support in the framework of the PRODEX Programme.
D.B. is supported by Spanish MCIN/AEI/10.13039/501100011033 grant PID2019-107061GB-C61 and No. MDM-2017-0737. 
L.D. acknowledges funding from the KU Leuven Interdisciplinary Grant (IDN/19/028), the European Union H2020-MSCA-ITN-2019 under Grant no. 860470 (CHAMELEON) and the FWO research grant G086217N.
I.K. acknowledge support from grant TOP-1 614.001.751 from the Dutch Research Council (NWO).
MPIA acknowledges support from the Federal Ministry of Economy (BMWi) through the German Space Agency (DLR).
J.P. acknowledges financial support from the UK Science and Technology Facilities Council, and the UK Space Agency.
G.O. acknowledge support from the Swedish National Space Board and the Knut and Alice Wallenberg Foundation.
P.P. thanks the Swiss National Science Foundation (SNSF) for financial support under grant number 200020\_200399.
P.R. thanks the European Space Agency (ESA) and the Belgian Federal Science Policy Office (BELSPO) for their support in the framework of the PRODEX Programme.
E. v D acknowledges support from A-ERC grant 101019751 MOLDISK.
The Cosmic Dawn Center (DAWN) is funded by the Danish National Research Foundation under grant No. 140. TRG is grateful for support from the Carlsberg Foundation via grant No. CF20-0534.
T.P.R acknowledges support from the ERC 743029 EASY. 
Support from SNSA is acknowledged.

\end{acknowledgement}

\bibliographystyle{aa}
\bibliography{main}

\begin{appendix}
\section{Stellar subtraction}
\label{app:other_red}
This appendix presents supplementary stellar subtraction methods, 
(traditional methods such as PCA and linear combination of the references observations). However, they do not effectively identify the planet, because of the inner disk that contributes to the mid-IR flux. 
\begin{figure}[H]
    \centering
    \includegraphics[width=9cm]{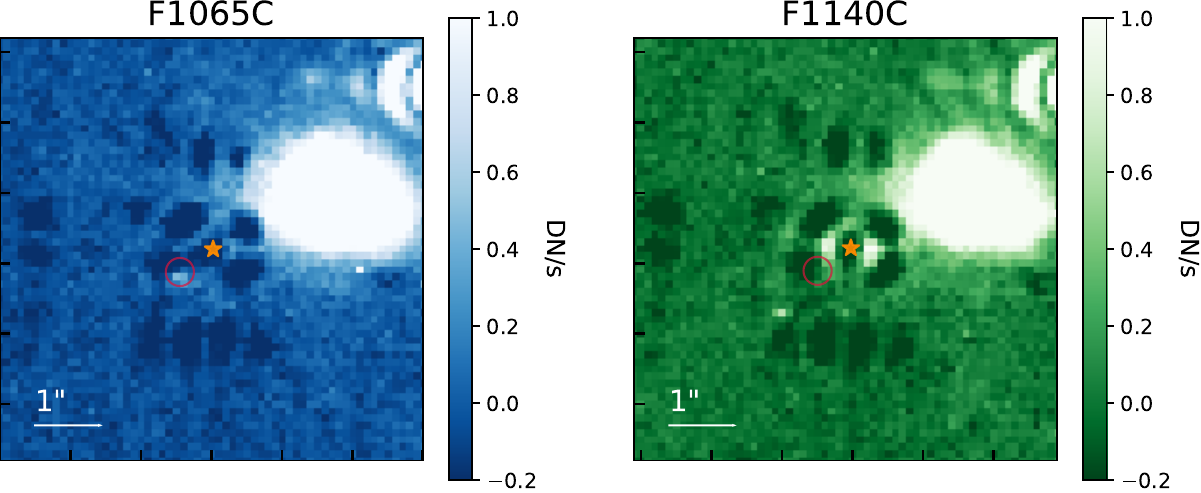}
    \includegraphics[width=9cm]{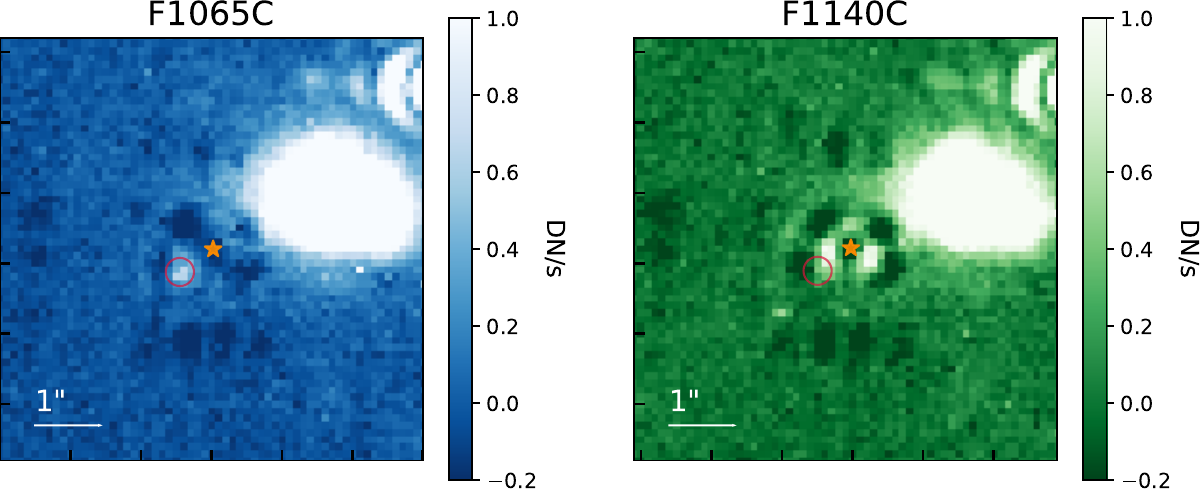}
    \caption{Subtraction of a reference image constructed by linear combination of the 9 references, without masking background objects (top) and by masking them (bottom). The expected position of the planet is shown in red, and the center of the coronagraph by the star in orange.}
    \label{fig:HD95_opt_lin}
\end{figure}

\begin{figure}[H]
    \centering
    \includegraphics[width=9cm]{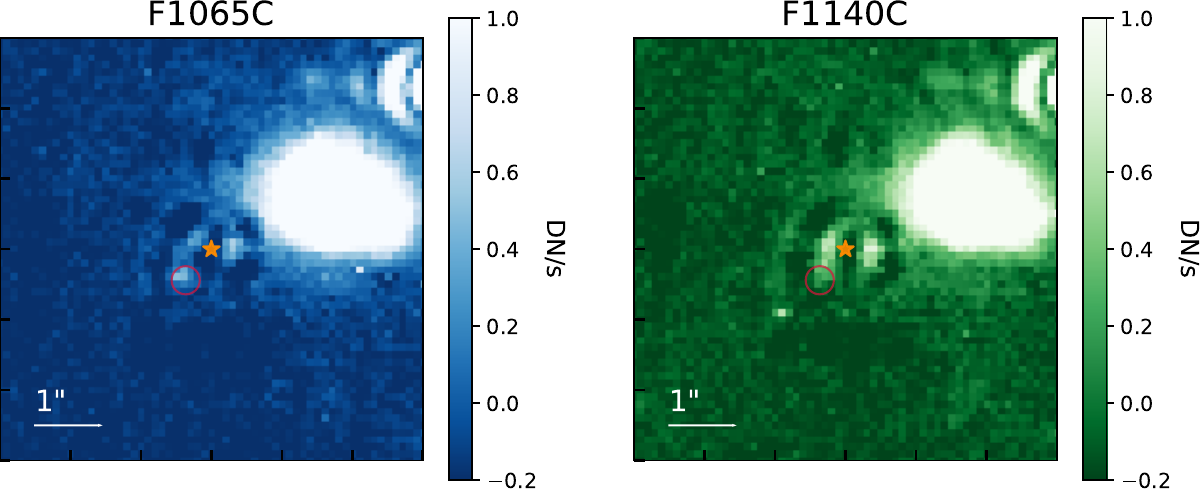}
    \caption{Subtraction of a reference after removing 9 PCA components. The expected position of the planet is shown in red and the center of the coronagraph by the star in orange.}
    \label{fig:HD95_PCA}
\end{figure}

\begin{figure}[H]
    \centering
    \includegraphics[width=5cm]{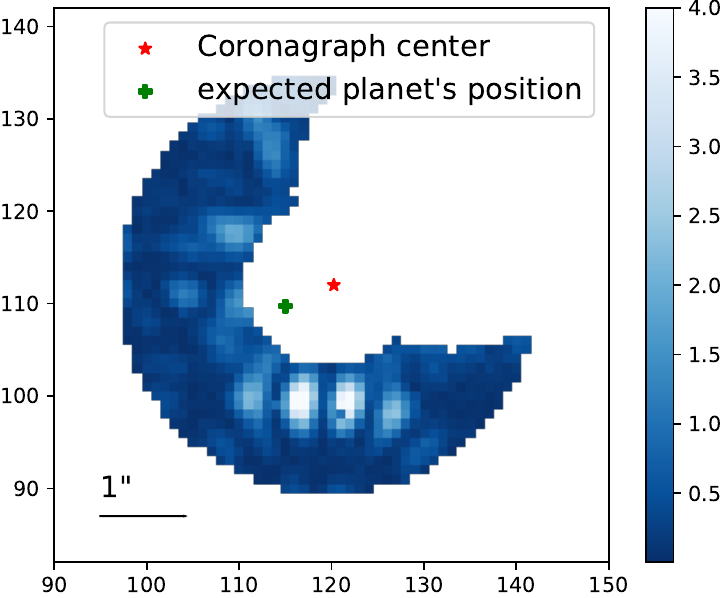}
    \caption{Region in which residuals are minimized, e.g. on the coronagraphic image in the F1065 filter.}
    \label{fig:mask_region_min_F1065}
\end{figure}

\begin{figure}[H]
    \centering
    \includegraphics[width=9cm]{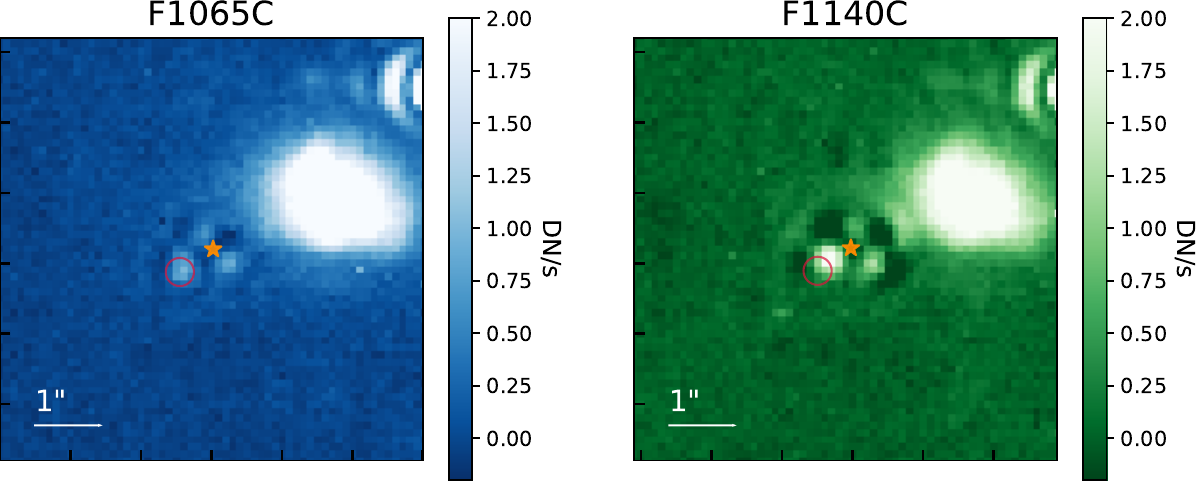}
    \caption{Subtraction of a reference constructed by linear combination of the 9 references, using the mask shown in Fig. \ref{fig:mask_region_min_F1065}.}
    \label{fig:HD95_many_ref_scale_opt_mask2}
\end{figure}

\begin{figure}[H]
    \centering
    \includegraphics[width=9cm]{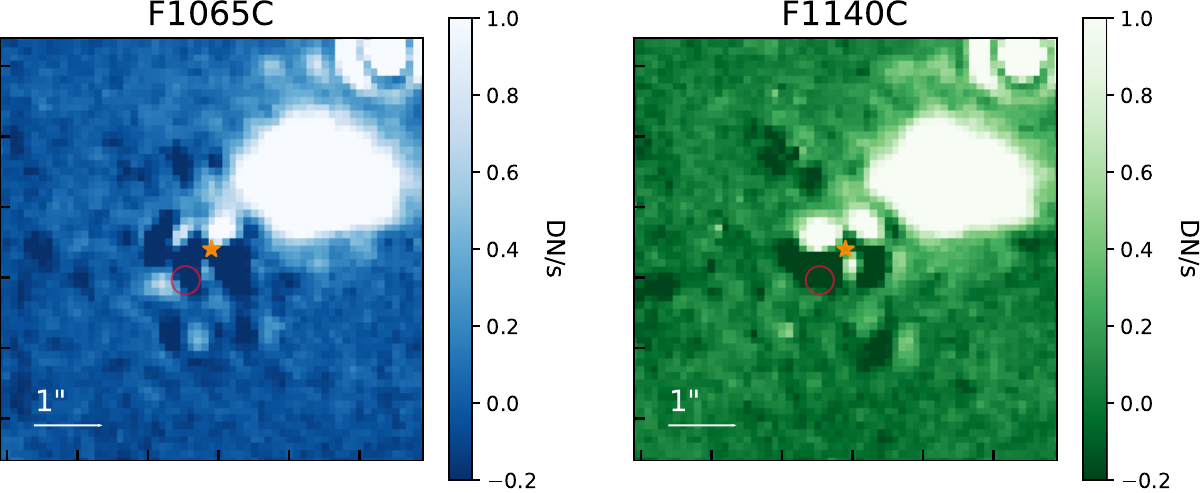}
    \caption{Subtraction of a reference constructed by linear combination of the library references, as detailed in Sect. \ref{sec:stellar_subtraction}.}
    \label{fig:HD95_lib_opt_lin}
\end{figure}

\section{Atmospheric fitting adding a blackbody component (CPD)}

\begin{figure*}[h!]
    \centering
    \includegraphics[width=18cm]{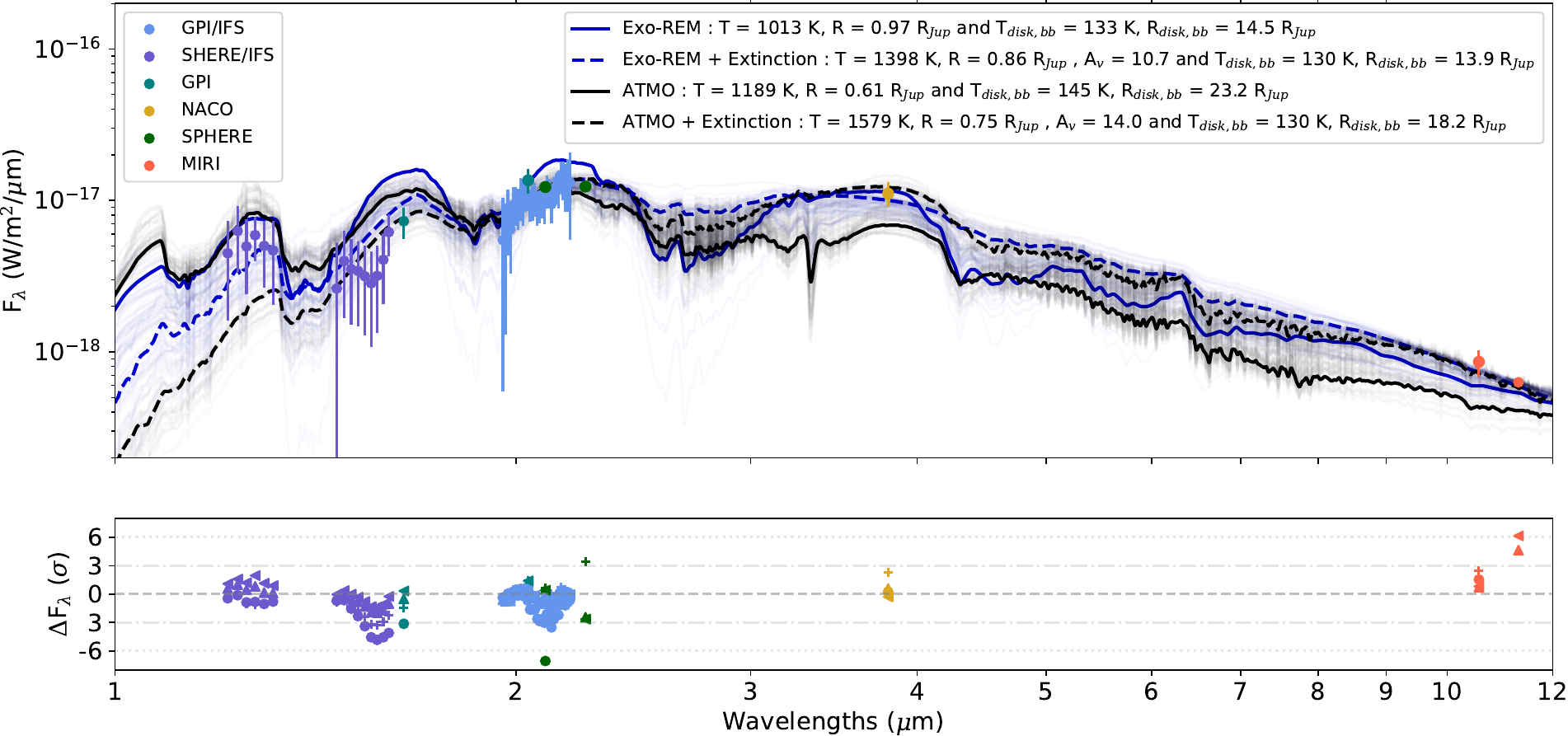}
    \caption{\texttt{Exo-REM} and \texttt{ATMO} best fits adding a CPD component in the fitting process, with all free priors. 
    The models with extinction are represented with dashed lines and those without extinction in plain lines. 
    The residuals between data and \texttt{Exo-REM} models without extinction are indicated with filled circles, and with crosses for \texttt{ATMO} models.
    The cases with extinction are indicated with up triangles for \texttt{Exo-REM} and left triangles for ATMO.}
    \label{fig:Spectra_FM_CPD_HD95086}
\end{figure*}

\begin{table*}[h!]
    \caption{Summary of the best-fit parameter corresponding to Fig. \ref{fig:Spectra_FM_CPD_HD95086}.}
    \centering
    \begin{tabular}{c|ccccccccccc}
        \hline
        \hline
         Model & T$_{eff}$ & log(g) & [Fe/H] & C/O & R (R$_{jup}$)& log(L) & M (M$_{Jup}$) & T$_{disk}$ & R$_{disk}$ (R$_{jup}$) & A$_v$\\
         \hline
         \hline
         \texttt{Exo-REM} 
         & 1013$^{+88}_{-67}$ & 
         3.35$^{+0.55}_{-0.25}$ & 0.55$^{+0.29}_{-0.35}$ & 0.37$^{+0.26}_{-0.21}$ & 0.97$^{+0.13}_{-0.18}$ & -4.98$^{+0.08}_{-0.06}$& 0.8$^{+2}_{0.4}$ & 133$^{+70}_{-26}$ & 14.5$^{+26.9}_{-11.4}$ & -- \\

         & 1398$^{+302}_{-236}$ & 
         4.07$^{+0.61}_{-0.71}$ & 0.21$^{+0.55}_{-0.49}$ & 0.45$^{+0.23}_{-0.24}$ & 0.86$^{+0.07}_{-0.09}$ & -4.56$^{+0.27}_{-0.25}$& 3.4$^{+10.0}_{-2.7}$ & 130$^{+47}_{-23}$ & 13.8$^{+23.0}_{-9.9}$ & 10.7$^{+5.2}_{-4.9}$\\

         \hline
         \texttt{ATMO} 
         & 1189$^{+68}_{-106}$ & 3.23$^{+0.52}_{-0.18}$ & 0.26$^{+0.18}_{-0.37}$ & 0.37$^{+0.11}_{-0.05}$ & 
         0.61$^{+0.21}_{-0.08}$ & 
         -4.96$^{+0.10}_{-0.07}$& 0.35$^{+0.43}_{-0.16}$ & 144$^{+391}_{-35}$ & 23.2$^{+39.6}_{-21.7}$\\

         & 1579$^{+273}_{-226}$ & 4.19$^{+0.75}_{-0.76}$ & -0.09$^{+0.34}_{-0.28}$ & 0.50$^{+0.13}_{-0.13}$ &
         0.75$^{+0.09}_{-0.12}$ &
         -4.47$^{+0.19}_{-0.19}$& 3.3$^{+16.1}_{-2.7}$ & 130$^{+45}_{-22}$ & 18.2$^{+22.5}_{-12.5}$ & 14.0$^{+3.8}_{-4.3}$\\

        \hline
    \end{tabular}
    \tablefoot{There are no priors on the atmospherics parameters, and we display both models with and without the extinction parameter A$_V$ taken into account.}
    \label{tab:models_best_fits_CPD}
\end{table*}

\clearpage
\section{Atmospheric fitting with only the near-infrared data}

\begin{figure*}[h!]
    \centering
    \includegraphics[width=18cm]{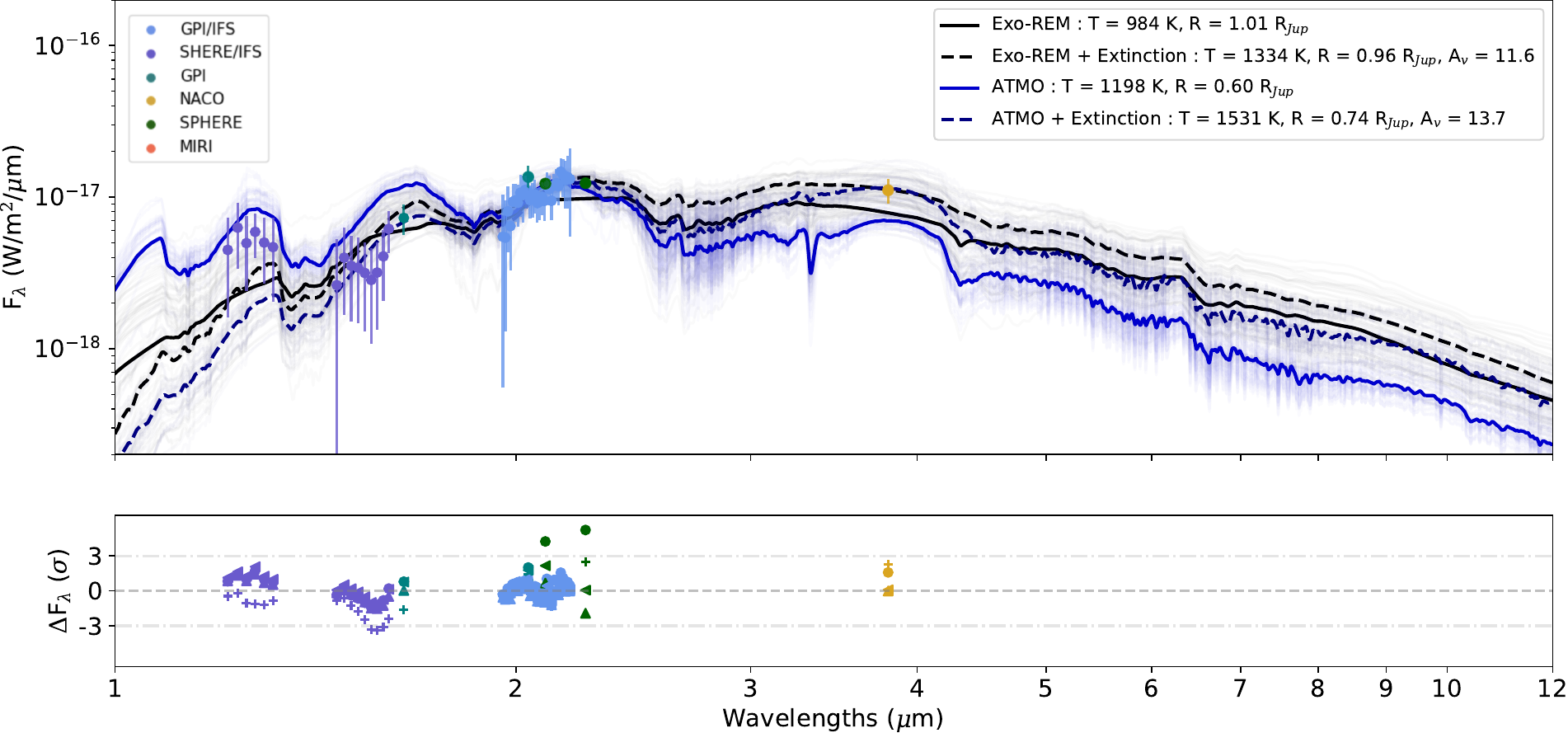}
    \caption{\texttt{Exo-REM} and \texttt{ATMO} best-fits models fitting only the near-IR points. 
    The models with extinction are represented with dashed lines and those without extinction in plain lines. 
    The residuals between data and \texttt{Exo-REM} models without extinction are indicated with filled circles, and with crosses for \texttt{ATMO} models. 
    The cases with extinction are indicated with up triangles for \texttt{Exo-REM} and left triangles for ATMO.}
    \label{fig:Spectra_FM_noMIRI_HD95086}
\end{figure*}

\begin{table*}[h!]
    \caption{Summary of the best-fit parameter corresponding to Fig. \ref{fig:Spectra_FM_noMIRI_HD95086}.}
    \centering
    \begin{tabular}{c|ccccccccccc}
        \hline
        \hline
         Model & T$_{eff}$ & log(g) & [Fe/H] & C/O & $\gamma$ & R (R$_{jup}$)& A$_v$ & log(L) & M (M$_{Jup}$)\\
         \hline
         \hline
         \texttt{Exo-REM} & \\
        free & 984+$^{+126}_{-122}$ & 3.33$^{+0.46}_{-0.22}$ & 0.49$^{+0.28}_{-0.39}$ & 0.28$^{+0.27}_{-0.12}$ & -- & 1.01$^{+0.51}_{-0.25}$ & -- & -5.03$^{+0.13}_{-0.08}$& 1.2$^{+1.9}_{-0.7}$\\

        prior g & 990+$^{+123}_{-118}$ & 3.90$^{+0.1}_{-0.1}$ & 0.53$^{+0.24}_{-0.27}$ & 0.34$^{+0.28}_{-0.16}$ & -- & 0.98$^{+0.35}_{-0.24}$ & -- & -5.06$^{+0.09}_{-0.07}$& 3.0$^{+2.8}_{-1.4}$\\

        prior R & 900+$^{+34}_{-42}$ & 3.38$^{+0.48}_{-0.25}$ & 0.51$^{+0.27}_{-0.32}$ & 0.26$^{+0.12}_{-0.10}$ & -- & 1.35$^{+0.05}_{-0.04}$ & -- & -4.95$^{+0.06}_{-0.08}$& 1.8$^{+3.5}_{-0.8}$\\

        \hline 
        free & 1334$^{+241}_{-179}$ & 4.11$^{+0.58}_{-0.68}$ & 0.20$^{+0.51}_{-0.47}$ & 0.47$^{+0.22}_{-0.23}$ & -- & 0.96$^{+0.18}_{-0.17}$ & 11.6$^{+4.2}_{-4.6}$ & -4.54$^{+0.22}_{-0.23}$ & 4.7$^{+13.4}_{-3.7}$\\

        prior g & 1329$^{+257}_{-202}$ & 3.90$^{+0.1}_{-0.1}$ & 0.18$^{+0.54}_{-0.46}$ & 0.48$^{+0.21}_{-0.25}$ & -- & 0.95$^{+0.19}_{-0.17}$ & 11.1$^{+4.5}_{-4.8}$ & -4.56$^{+0.22}_{-0.23}$ & 2.9$^{+1.3}_{-0.9}$\\

        prior R & 1092$^{+199}_{-129}$ & 3.98$^{+0.66}_{-0.63}$ & 0.38$^{+0.39}_{-0.61}$ & 0.49$^{+0.21}_{-0.25}$ & -- & 
        1.33$^{+0.05}_{-0.05}$ & 10.7$^{+5.4}_{-6.0}$ & -4.62$^{+0.28}_{-0.21}$ & 6.8$^{+24.0}_{-5.2}$\\
        
         \hline 
         \hline 
         \texttt{ATMO} & \\
               
        free & 1198+$^{+59}_{-70}$ & 3.32$^{+0.57}_{-0.23}$ & 0.30$^{+0.14}_{-0.35}$ & 0.38$^{+0.11}_{-0.05}$ & 1.03$^{+0.01}_{-0.01}$ & 0.60$^{+0.10}_{-0.06}$ & -- & -5.14$^{+0.05}_{-0.05}$ & 0.4$^{+0.8}_{-0.2}$\\

        prior g & 1206+$^{+55}_{-53}$ & 3.89$^{+0.1}_{-0.1}$ & 0.31$^{+0.13}_{-0.26}$ & 0.40$^{+0.13}_{-0.07}$ & 1.03$^{+0.01}_{-0.0}$ & 0.58$^{+0.06}_{-0.05}$ & -- & -5.17$^{+0.04}_{-0.04}$ & 1.1$^{+0.4}_{-0.3}$\\

        prior R & 931+$^{+16}_{-16}$ & 3.08$^{+0.14}_{-0.06}$ & -0.37$^{+0.19}_{-0.09}$ & 0.35$^{+0.07}_{-0.03}$ & 1.02$^{+0.01}_{-0.0}$ & 1.35$^{+0.05}_{-0.05}$ & -- & -4.89$^{+0.03}_{-0.03}$ & 0.9$^{+0.3}_{-0.1}$\\

        \hline
        free & 1531$^{+319}_{-233}$ & 4.17$^{+0.78}_{-0.73}$ & -0.10$^{+0.33}_{-0.26}$ & 0.50$^{+0.13}_{-0.13}$ & 1.03$^{+0.01}_{-0.01}$ & 0.74$^{+0.20}_{-0.14}$ & 13.7$^{+3.7}_{-4.0}$ & -4.51$^{+0.19}_{-0.18}$ & 3.5$^{+16.6}_{-2.8}$\\

        prior g & 1493$^{+322}_{-210}$ & 3.90$^{+0.1}_{-0.1}$ & -0.09$^{+0.36}_{-0.27}$ & 0.50$^{+0.13}_{-0.13}$ & 1.03$^{+0.01}_{-0.01}$ & 0.74$^{+0.20}_{-0.14}$ & 12.8$^{+4.1}_{-3.8}$ & -4.55$^{+0.20}_{-0.18}$ & 1.8$^{+1.2}_{-0.7}$\\

        prior R & 1192$^{+67}_{-63}$ & 3.91$^{+0.75}_{-0.67}$ & 0.03$^{+0.35}_{-0.36}$ & 0.48$^{+0.15}_{-0.12}$ & 1.03$^{+0.01}_{-0.01}$ & 1.33$^{+0.05}_{-0.05}$ & 13.6$^{+2.4}_{-2.3}$ & -4.47$^{+0.1}_{-0.1}$ & 5.7$^{+26.6}_{-2.5}$\\
        
        \hline
         
    \end{tabular}
    \tablefoot{Same as Tab. \ref{tab:models_best_fits} using only the near-IR data.}
    \label{tab:models_best_fits_nomiri}
\end{table*}

\begin{figure*}[h!]
    \centering
    \includegraphics[width=18cm]{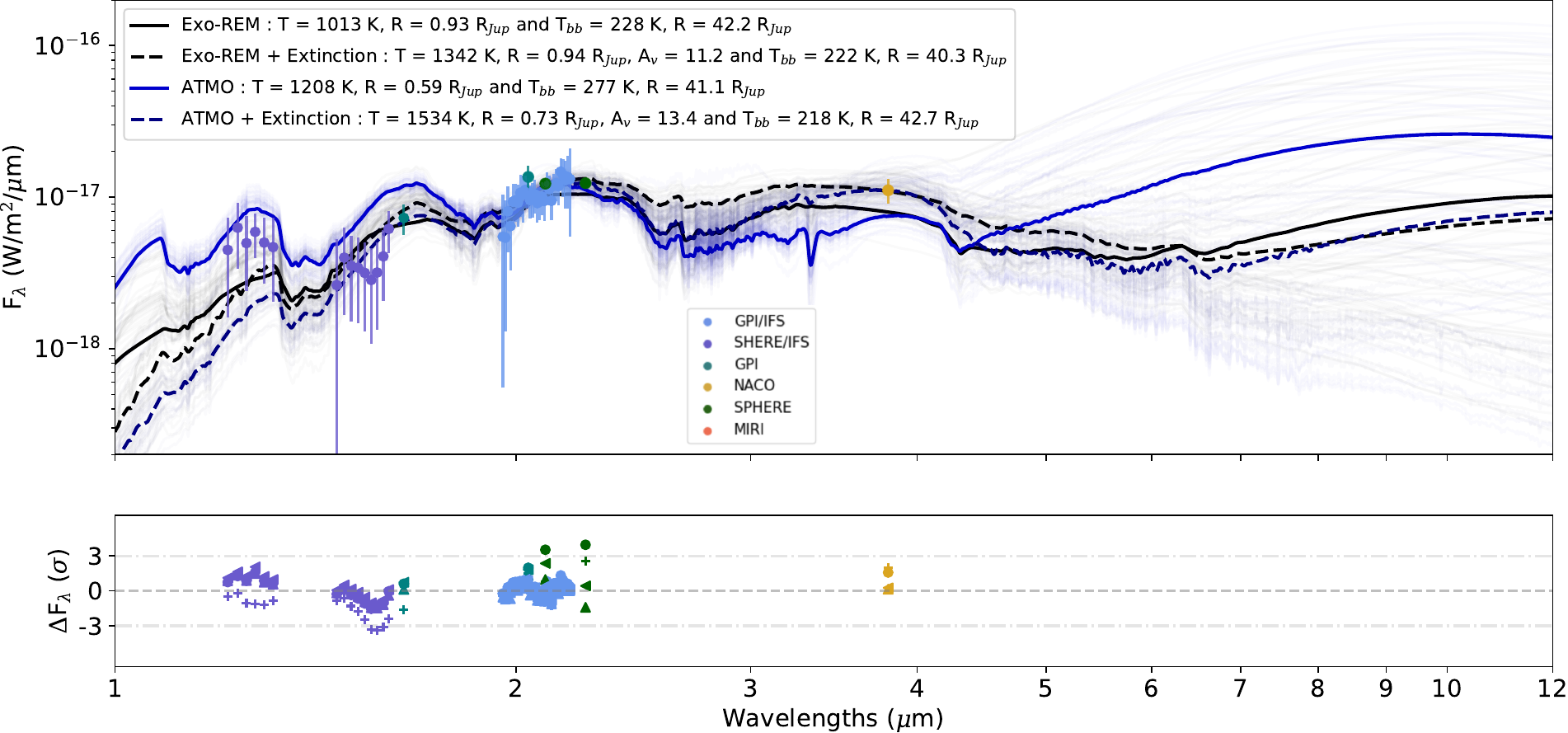}
    \caption{\texttt{Exo-REM} and \texttt{ATMO} best-fits models, fitting only the near-IR points and adding a blackbody component to account for a CPD contribution.
    The models with extinction are represented with dashed lines and those without extinction in plain lines. 
    The residuals between data and \texttt{Exo-REM} models without extinction are indicated with filled circles, and with crosses for \texttt{ATMO} models.
    The cases with extinction are indicated with up triangles for \texttt{Exo-REM} and left triangles for ATMO.}
    \label{fig:Spectra_FM_noMIRI_CPD_HD95086}
\end{figure*}

\begin{table*}[h!]
    \caption{Summary of the best-fit parameter for the near-IR data adding a CPD component.}
    \centering
    \begin{tabular}{c|ccccccccccc}
        \hline
        \hline
         Model & T$_{eff}$ & log(g) & [Fe/H] & C/O & R (R$_{jup}$)& log(L) & M (M$_{Jup}$) & T$_{disk}$ & R$_{disk}$ (R$_{jup}$) & A$_v$\\
         \hline
         \hline
         \texttt{Exo-REM} 
         & 1013$^{113}_{-130}$ & 
         3.36$^{+0.50}_{-0.24}$ & 0.51$^{+0.28}_{-0.38}$ & 0.31$^{+0.27}_{-0.14}$ & 0.93$^{+0.47}_{-0.21}$ & -4.38$^{+0.59}_{-0.46}$& 1.0$^{+1.9}_{0.6}$ & 228$^{+82}_{-80}$ & 42.2$^{+36.7}_{-29.3}$ & -- \\

         & 1342$^{+242}_{-184}$ & 
         4.11$^{+0.57}_{-0.68}$ & 0.21$^{+0.52}_{-0.46}$ & 0.47$^{+0.21}_{-0.23}$ & 0.94$^{+0.18}_{-0.16}$ & -4.21$^{+0.42}_{-0.31}$& 4.5$^{+12.9}_{3.5}$ & 222$^{+83}_{-76}$ & 40.3$^{+38.0}_{-28.2}$ & 11.2$^{+4.5}_{-4.6}$\\

         \hline
         \texttt{ATMO} 
         & 1208$^{+56}_{-61}$ & 3.34$^{+0.53}_{-0.25}$ & 0.29$^{+0.15}_{-0.30}$ & 0.39$^{+0.11}_{-0.06}$ & 0.59$^{+0.11}_{-0.06}$ & -4.16$^{+0.61}_{-0.64}$& 0.3$^{+0.7}_{-0.1}$ & 277$^{+76}_{-99}$ & 41.1$^{+38.2}_{-28.8}$ & --\\
      
         & 1534$^{+285}_{-221}$ & 4.14$^{+0.75}_{-0.70}$ & -0.08$^{+0.34}_{-0.27}$ & 0.50$^{+0.12}_{-0.13}$ & 0.73$^{+0.19}_{-0.13}$ & -4.48$^{+0.18}_{-0.17}$& 3.1$^{+14.7}_{-2.5}$ & 218$^{+80}_{-74}$ & 42.7$^{+36.1}_{-29.3}$ & 13.4$^{+3.6}_{-4.0}$\\

        \hline
    \end{tabular}
    \tablefoot{Without any priors, with and without the extinction parameter A$_V$. Same as Tab. \ref{tab:models_best_fits_CPD} using only the near-IR data.}
    \label{tab:models_best_fits_CPD_noMIRI}
\end{table*}

\clearpage

\section{Suspected new point source}
\begin{figure*}[h]
    \centering
    \includegraphics[width=14cm]{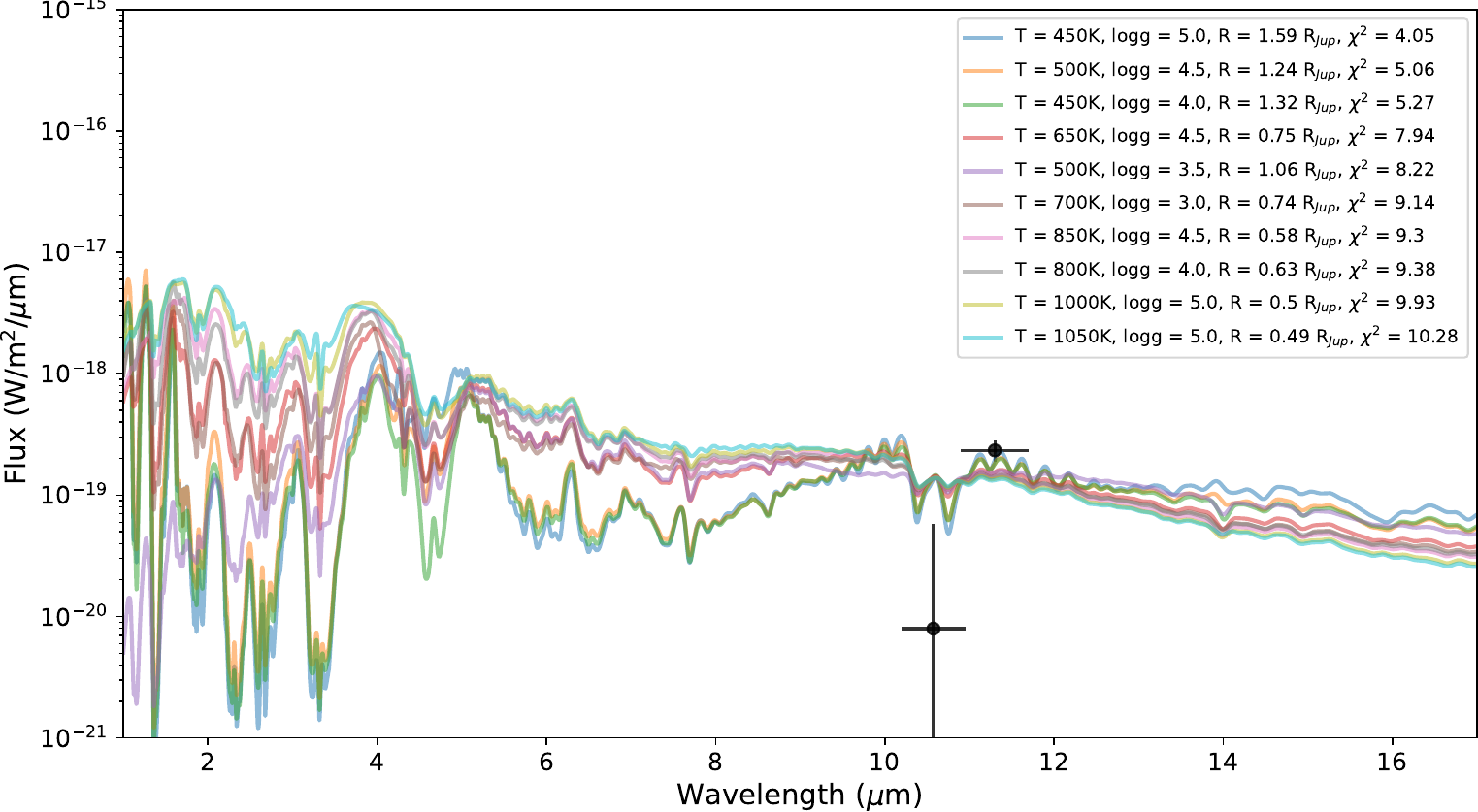}
    \caption{Photometry extracted from the model presented in Fig. \ref{fig:model_new_source} (black points). \texttt{Exo-REM} photometric models with the minimum $\chi^2$ are also represented (based only on these two points).}
    \label{fig:photom_new_source}
\end{figure*}
\begin{figure*}[h]
    \centering
    \includegraphics[width=18cm]{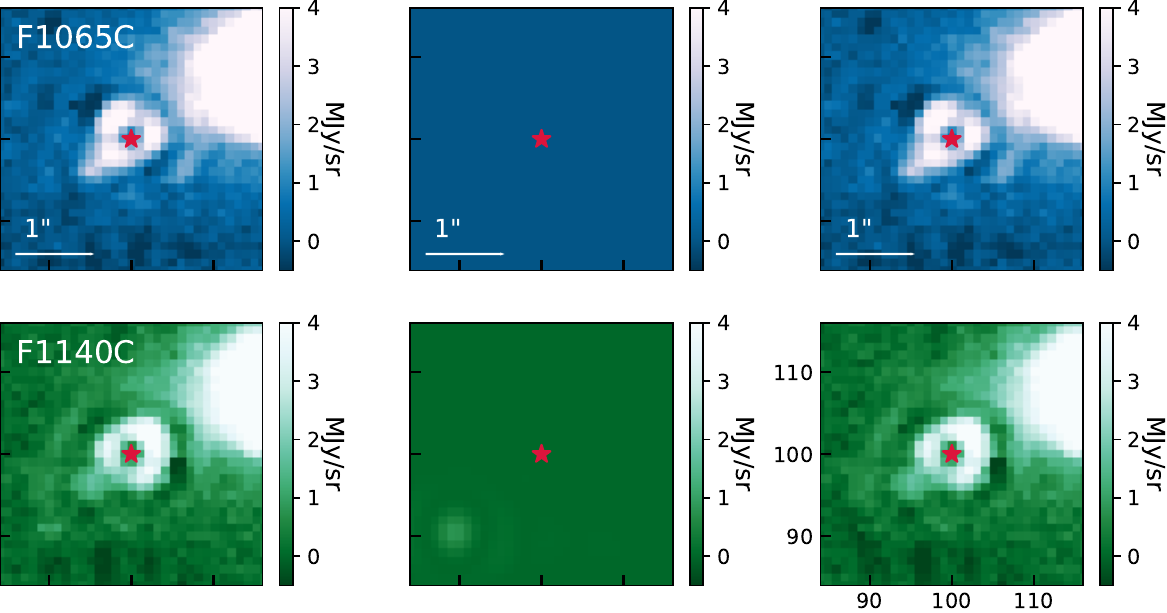}
    \caption{Modeling of the few bright pixels observed in the SE in the F1140C. Left : Data from both filters, Middle : \texttt{WebbPSF} Models, and right : residuals after subtraction.}
    \label{fig:model_new_source}
\end{figure*}
\end{appendix}

%
%
\end{document}